\documentclass{emulateapj}
\usepackage{apjfonts}
\input{epsf}
 
\slugcomment{To appear in The Astronomical Journal}

\shorttitle{Spectroscopic catalog of the brightest, northern M dwarfs}
\shortauthors{L\'epine et al.}

\begin{document}

\title{A Spectroscopic Catalog of the Brightest ($J<9$) M Dwarfs in
  the Northern Sky\altaffilmark{*,$\dagger$}}

\author{S\'ebastien L\'epine\altaffilmark{1,2}, Eric J.
  Hilton\altaffilmark{3,4}, Andrew W. Mann\altaffilmark{3}, Matthew
  Wilde\altaffilmark{1}, B\'arbara Rojas-Ayala\altaffilmark{1}, Kelle
  L. Cruz\altaffilmark{1,2,5}, \& Eric Gaidos\altaffilmark{4}}

\altaffiltext{*}{Based on observations collected at the MDM
  Observatory, located on Kitt Peak, and operated jointly by the
  University of Michigan, Dartmouth College, the Ohio State
  University, Columbia University, and the University of Ohio.}

\altaffiltext{$\dagger$}{Based on observations collected at the
  University of Hawaii 2.2-meter telescope, located on Mauna Kea, and
  operated by the University of Hawaii.}

\affil{$^1$ Department of Astrophysics, American Museum of
  Natural History, Central Park West at 79th Street, New York, NY
  10024, USA; lepine@amnh.org, brojas-ayala@amnh.org, mwilde@amnh.org}

\affil{$^2$ Department of Physics, Graduate Center, City
  University of New York, 365 Fifth Avenue New York, NY 10016, USA}

\affil{$^3$ Institute for Astrophysics, University of
  Hawai'i, Honolulu, HI 96822, USA; hilton@ifa.hawaii.edu,
  amann@ifa.hawaii.edu}

\affil{$^4$ Department of Geology \& Geophysics, University of
  Hawai'i, 1680 East-West Road, Honolulu, HI 96822, USA;
  gaidos@hawaii.edu}

\affil{$^5$ Department of Physics and Astronomy, Hunter College,
  City University of New York, 695 Park Avenue, New York, NY 10065,
  kellecruz@gmail.com}

\begin{abstract}
We present a spectroscopic catalog of the 1,564 brightest ($J<9$) M
dwarf candidates in the northern sky, as selected from the SUPERBLINK
proper motion catalog. Observations confirm 1,408 of the candidates to
be late-K and M dwarfs with spectral subtypes K7-M6. From the low
($\mu>$40 mas yr$^{-1}$) proper motion limit and high level of
completeness of the SUPERBLINK catalog in that magnitude range, we
estimate that our spectroscopic census most likely includes $>90\%$ of
all existing, northern-sky M dwarfs with apparent magnitude $J<9$. Only
682 stars in our sample are listed in the Third Catalog of Nearby
Stars (CNS3); most others are relative unknowns and have spectroscopic
data presented here for the first time. Spectral subtypes are assigned
based on spectral index measurements of CaH and TiO molecular bands; a
comparison of spectra from the same stars obtained at different
observatories however reveals that spectral band index measurements are
dependent on spectral resolution, spectrophotometric calibration, and
other instrumental factors. As a result, we find that a consistent
classification scheme requires that spectral indices be calibrated and
corrected for each observatory/instrument used. After systematic
corrections and a recalibration of the subtype-index relationships for
the CaH2, CaH3, TiO5, and TiO6 spectral indices, we find that we can
consistently and reliably classify all our stars to a half-subtype
precision. The use of corrected spectral indices further requires us
to recalibrate the $\zeta$ parameter, a metallicity indicator based on
the ratio of TiO and CaH optical bandheads. However, we find that our
$\zeta$ values are not sensistive enough to diagnose metallicity
variations in dwarfs of subtypes M2 and earlier ($\pm0.5$dex accuracy)
and are only marginally useful at later M3-M5 subtypes ($\pm0.2$dex
accuracy). Fits of our spectra to the Phoenix atmospheric model grid
are used to estimate effective temperatures. These suggest the
existence of a plateau in the M1-M3 subtype range, in agreement with
model fits of infrared spectra but at odds with photometric
determinations of $T_{eff}$. Existing geometric parallax measurements
are extracted from the literature for 624 stars, and are used to
determine spectroscopic and photometric distances for all the other
stars. Active dwarfs are identified from measurements of H$\alpha$
equivalent widths, and we find a strong correlation between H$\alpha$
emission in M dwarfs and detected X-ray emission from ROSAT and/or a
large UV excess in the GALEX point source catalog. We combine proper
motion data and photometric distances to evaluate the (U,V,W)
distribution in velocity space, which is found to correlate tighly
with the velocity distribution of G dwarfs in the Solar
Neighborhood. However, active stars show a smaller dispersion in their
space velocities, which is consistent with those stars being younger
on average. Our catalog will be most useful to guide the selection of
the best M dwarf targets for exoplanet searches, in particular those
using high-precision radial velocity measurements.
\end{abstract}


\section{Introduction}


M dwarfs have become targets of choice for many exoplanet surveys. This
is because low-mass planets (i.e. Earth- to Neptune-size) are easier
to detect by the Doppler or transit techniques around stars of lower
mass. The transits of smaller planets are also easier to detect when
they occur in the also smaller M dwarfs. In addition, M dwarfs have
much lower luminosities than the Sun, and their ``habitable zones''
(HZ) are closer in, which makes transits more likely to occur and
radial velocity variations easier to detect for planets in their
HZ. Earth-like planets within the HZ of M dwarfs are thus eminently
more detectable with current observational techniques than earth-like
planets in the HZ of G dwarfs \citep{Tarter2007,Gaidos2007}. M dwarfs
are also the most plentiful class of stars, constituting the largest
fraction ($>70\%$) of main sequence objects in the Galaxy and in the
vicinity of the Sun \citep{Reid2002,Covey2008,Bochanski2010}. 

However, even nearby M dwarfs are generaly faint at the visible
wavelengths where most planet searches are conducted, and most
exoplanet detection techniques \--- with the notable exception of
micro-lensing \citep{Dong_etal.2009} \--- are currently restricted
to relatively bright stars. This significantly limits the number of M
dwarfs that can be targeted in exoplanet surveys. Doppler searches in
particular are usually restricted to stars with visual magnitudes
$V<12$, and less than $\sim$10\% of late-K and early-M stars within
30~pc are currently being monitored by the large-scale Doppler surveys
\citep{Butler2006,Mayor2009}. However, new surveys are pushing this
limit to fainter magnitudes \citep{Apps2009}, and high-resolution
spectrographs suitable for Doppler observations at near-infrared
wavelengths, where M dwarfs are relatively brighter, are being
developed \citep{Terada2008,Bean2010,Quirrenbach2010,Wang2010}. In any
case, only a fraction of all catalogued, nearby M dwarfs are bright
enough to be included in radial-velocity monitoring programs. 

Transit surveys, on the other hand, can include much fainter
stars \citep{Irwin2009}. However, because they have a much lower
detection efficiency due to orbital inclination constraints, they
require extensive lists (thousands) of targets in order to detect any
significant number of transit events. For transit surveys, the Solar
Neighborhood census and its estimated $\approx$5,000 M dwarfs is
therefore too small, and transit programs would greatly benefit from
extending their target lists to much larger distance limits.

A fundamental obstacle to progress has been the lack of a large,
complete, and uniform catalog of bright M dwarfs suitable as targets
for exoplanet programs. In particular, most catalogs and surveys of M
dwarfs have focused on identifying the nearest objects, which are not
necessarily the brightest. Whereas the Hipparcos catalog
\citep{vanLeeuwen2007} provides a near-complete census of solar-mass
stars within 100 parsecs of the Sun, the bright magnitude limit of the
catalog excludes all but the very nearest M dwarfs \--- although it
lists stars as faint as $V=12-13$, the Hipparcos catalog is complete
only to $V<8$.

The widely utilized {\it Third Catalog of Nearby Stars}, or CNS3
\citep{gj1991}, which lists $\approx$3,800 stars, though predating the
Hipparcos survey, has historically provided a more complete list of
M dwarf candidates in the Solar Neighborhood. Many of the fainter
stars in the CNS3 have (ground-based) parallax measurements from a
variety of sources \citep{VLH95}. However, the CNS3 was largely
compiled based on a photometric analysis the high proper motion stars
catalogued by \citet{lhs,nltt}, and in large part using photometric
data collected by \citet{Gliese_1982} and
\citet{Weis1984,Weis1986,Weis1987}. The CNS3 has been largely used in
recent years to select M dwarf targets for exoplanet surveys
\citep{Marcy2001,Naef2003,Butler2004,Rivera2005,Endl2008,Bailey2009,Anglada2012a,Anglada2012b}. Unfortunately,
the catalog suffers from various sources of incompleteness. These
mainly consist of: (1) limited availability of quality data for stars
in the Luyten catalogs, at the time the CNS3 was compiled, (2)
kinematic bias in the Luyten catalog due a relatively high
($\mu>0.18\arcsec$ yr$^{-1}$) proper motion limit at the low end, and
(3) incompleteness of the Luyten catalogs even for stars with proper
motions above the fiducial limit.

Motivated mainly by the need to complete the census of the Solar
Neighborhood, several surveys have since been conducted to identify
the low-mass stars (mostly M dwarfs) suspected to be missing from the
CNS3. These have included a re-analysis of the proper motion catalogs
of \citet{nltt,lhs} in light of high quality photometric data
provided by the 2MASS survey \citep{Cutri.etal.2003}. This has led to
the identification of hundreds of additional nearby star candidates
that had previously been overlooked
\citep{Reid_Cruz.2002,Reidetal.2004}. In addition, cross-matching
against 2MASS and examination of Digitized Sky Survey images has
uncovered significant ($>1\arcmin$) errors in many of the coordinates
quoted in the Luyten catalogs, which was hitherto preventing efficient
follow-up studies \citet{Bakos.2002,SalimGould.2003}. 

Parallelling these efforts, new proper motion surveys have been
conducted, mainly to find the high proper motion stars missing from
the Luyten catalogs with a focus on completing the stellar census of
the Solar Neighborhood
\citep{Lepine2002,Lepine2003b,Deacon2005,Levine2005,Lepine2005,Subasavage2005a,Subasavage2005b,Lepine2008}. In
addition, some surveys have also been reaching to lower proper motion
limits \citep{Lepine2005,Reid2007,Boyd2011}, potentially extending the
census of M dwarfs to larger distances. Recently, we have analyzed
data from theSUPERBLINK proper motion survey, which has a proper
motion limit $\mu>0.04\arcsec$ yr$^{-1}$, with an emphasis on the
identification of {\em bright} M dwarfs, rather than just {\em nearby}
ones; our search has turned up 8,889 candidate M dwarfs with infrared
magnitude $J<10$ \citep{LepineGaidos.2011}. Of these, we found that
only 982 were previously listed in the Hipparcos catalog, and another
898 in the CNS3. Most of the other 7009 stars were not commonly known
objects, and were identified as probable nearby M dwarfs for the first
time. With its high estimated completeness, especially in the northern
sky, the \citet{LepineGaidos.2011} census provides a solid basis for
assembling an extensive and highly complete catalog of bright M
dwarfs, suitable for exoplanet search programs.


Not all M dwarfs, however, are equally suitable targets for planet
searches. Some M dwarfs have significant photometric variability
(flares, spots) which are affecting transit searches
\citep{Hartman2011}; some display chromospheric emission affecting
Doppler searches \citep{Isaacson2010}. Because M dwarfs are relatively
faint stars, they often require considerable investement of observing
time on large telescopes to achieve exoplanet detection, and there is
value in identifying subsets of M dwarfs that are intrinsically more
likely to host detectable planets \citet{Herrero2011}. In
particular, one might be interested in selecting stars of higher
metallicity which may harbor more massive planets
\citep{Sousa_etal.2010}, or young stars with relatively luminous
massive planets which would be easier to detect through direct imaging
\citep{Mugrauer_etal.2010}. In addition, one would like to avoid
possible contaminants (e.g. background giants) or problematic systems
(e.g. very active stars) in order to optimize exoplanet survey
efficiencies. 

Determining physical properties of the M dwarfs is also important in
order to better characterize the local populations of low-mass
stars. This is especially true since proximity makes them brighter and
thus more efficient targets for follow-up observations and detailed
study. Some of the bright M dwarfs may be close enough
(d$\lesssim$20pc) to warrant inclusion in the parallax programs
devoted to completing the census of low-mass stars in the Solar
Neighborhood \citep{Henry_etal.2006}, in which case it is also
important that the candidates first be vetted through spectral
typing. 

\begin{deluxetable*}{llrrrrrrrrrcrrr}
\tabletypesize{\scriptsize}
\tablecolumns{15} 
\tablewidth{0pt} 
\tablecaption{Survey stars: positions and photometry. \label{table_photo}}
\tablehead{
\colhead{Star name} & 
\colhead{CNS3\tablenotemark{a}} & 
\colhead{R.A.(ICRS)} &
\colhead{Decl.(ICRS)} &
\colhead{$\mu_{R.A.}$} &
\colhead{$\mu_{Decl.}$} &
\colhead{Xray\tablenotemark{b}} &
\colhead{hr1\tablenotemark{b}} &
\colhead{FUV\tablenotemark{c}} &
\colhead{NUV\tablenotemark{c}} &
\colhead{V} &
\colhead{V} &
\colhead{J\tablenotemark{d}} &
\colhead{H\tablenotemark{d}} &
\colhead{K$_s$\tablenotemark{d}} \\
\colhead{} &
\colhead{} &
\colhead{(ICRS)} &
\colhead{(ICRS)} &
\colhead{$\arcsec$ yr$^{-1}$} &
\colhead{$\arcsec$ yr$^{-1}$} &
\colhead{cnts/s} &
\colhead{} &
\colhead{mag} &
\colhead{mag} &
\colhead{mag} &
\colhead{flag} &
\colhead{mag} &
\colhead{mag} &
\colhead{mag} 
}
\startdata  
PM I00006+1829  &            &    0.163528 & 18.488850 & 0.335 & 0.195&       &      &      & 20.04& 11.28&T&  8.44&  7.79&  7.64\\
PM I00012+1358S &            &    0.303578 & 13.972055 & 0.025 & 0.144&       &      &      & 19.85& 11.12&T&  8.36&  7.71&  7.53\\
PM I00033+0441  &            &    0.829182 &  4.686940 &-0.024 &-0.085&       &      &      & 21.18& 12.04&T&  8.83&  8.18&  7.98\\
PM I00051+4547  & Gl  2      &    1.295195 & 45.786587 & 0.870 &-0.151&       &      &      &      &  9.95&T&  6.70&  6.10&  5.85\\
PM I00051+7406  &            &    1.275512 & 74.105217 & 0.035 &-0.023&       &      &      &      & 10.63&T&  7.75&  7.15&  6.97\\
PM I00077+6022  &            &    1.927582 & 60.381760 & 0.340 &-0.027& 0.1700& -0.41&      &      & 14.26&P&  8.91&  8.33&  8.05\\
PM I00078+6736  &            &    1.961682 & 67.607124 &-0.045 &-0.091&       &      &      &      & 12.18&P&  8.35&  7.72&  7.51\\
PM I00081+4757  &            &    2.026727 & 47.950695 &-0.119 & 0.003& 0.2190& -0.27& 19.68& 18.91& 12.70&P&  8.52&  8.00&  7.68\\
PM I00084+1725  & GJ 3008    &    2.113679 & 17.424309 &-0.093 &-0.064&       &      &      & 19.24& 10.73&T&  7.81&  7.16&  6.98\\
PM I00088+2050  & GJ 3010    &    2.224675 & 20.840403 &-0.065 &-0.247& 0.0899& -0.28& 21.07& 16.71& 13.90&P&  8.87&  8.26&  8.01\\
PM I00110+0512  &            &    2.769255 &  5.208822 & 0.241 & 0.061&       &      & 22.85& 20.58& 11.55&T&  8.53&  7.88&  7.69\\
PM I00113+5837  &            &    2.841032 & 58.617561 & 0.056 & 0.029&       &      &      &      & 11.21&T&  8.02&  7.31&  7.13\\
PM I00118+2259  &            &    2.970996 & 22.984573 & 0.142 &-0.221& 0.4110&  0.28&      & 22.37& 13.09&P&  8.86&  8.31&  7.99\\
PM I00125+2142En&            &    3.139604 & 21.713478 & 0.189 &-0.290&       &      &      &      & 11.67&T&  8.84&  8.28&  8.04\\
PM I00131+7023  &            &    3.298130 & 70.398003 & 0.045 & 0.139&       &      &      & 19.93& 11.37&T&  8.26&  7.59&  7.39
\enddata     
\tablenotetext{a}{Designation in the Third Catalog of Nearby Stars \citep{gj1991}}
\tablenotetext{b}{X-ray flux and hardness ratio from the ROSAT all-sky
  points source catalog \citep{Voges.etal.1999,Voges.etal.2000}.}
\tablenotetext{c}{Far-UV and near-UV magnitudes in the GALEX fifth data release.}
\tablenotetext{d}{Infrared magnitudes from the Two Micron All-Sky
  Survey \citep{Cutri.etal.2003}.}
\end{deluxetable*} 


Spectral classification and analysis for a significant fraction of the
low-mass stars in the CNS3 was performed as part of the Palomar-MSU
spectroscopic survey \citep{Reid1995,Hawley1996}, hereafter PMSU. The
survey has notably provided formal spectral classification for 1,971
of the fainter CNS3 stars, confirming 1,648 of them to be nearby M
dwarfs. More recent spectroscopic follow-up surveys have mainly
focused on candidate nearby stars missing from the CNS3. Very high
proper motion stars from the Luyten catalogs
\citep{GizisReid.1997,ReidGizis.2005}, or stars discovered in the more
recent proper motion surveys
\citet{Scholz2002,Lepine2003,Scholz2005,Reyle2006} have thus been
targeted. Most notably, the ``Meet the Cool Neighbours'' program
(hereafter MCN) has determined spectral subtypes for several hundred M
dwarfs identified from the Luyten catalogs but not listed in the CNS3
\citep{CruzReid.2002,Cruzetal.2003,Reidetal.2003,Reidetal.2004,Cruzetal.2007,Reid2007}. As
with the other more recent surveys, the MCN program placed an emphasis
on the identification and classification of nearby, very-cool M
dwarfs, most of which are however relatively faint and unsuitable for
exoplanet surveys. It should be noted that while large numbers of M
dwarfs have also been identified and classified as part of the
spectroscopic follow-up program of the Sloan Digital Sky Survey
\citep{Bochanski2005,Bochanski2010,West2011}, most of them are
relatively distant sources, and generally too faint for exoplanet
surveys.

This is why a significant fraction of the bright M dwarf candidates
published in \citet{LepineGaidos.2011} had no available spectroscopic
data at the time of release. In order to assemble a comprehensive
database of M dwarfs targets suitable for exoplanet survey programs,
we are now conducting a spectroscopic follow-up survey of the
brightest M dwarf candidates from \citep{LepineGaidos.2011}. Our goal
is to provide a uniform catalog of spectroscopic measurements to
confirm the M dwarf classification, and initiate detailed studies of
their physical properties, as well as the tayloring of exoplanet
searches. In this paper, we present the first results of our survey,
which provides data for the 1,564 brightest M dwarf candidates north
of the celestial equator, with apparent near-infrared magnitudes
J$<$9. Observations are described in \S2. Our spectral classification
techniques  are described in \S3, and our model fitting and effective
temperature determinations are given in \S4. Metallicity measurements
are presented in \S5. Activity diagnostics are presented and analyzed
in \S6. A kinematic study informed by our metallicity and activity
measurements is presented in \S7, followed by discussion and
conclusions in \S8.

\section{Spectroscopic observations}

\subsection{Target selection}

Targets for the follow-up spectroscopic program were selected from the
catalog of 8,889 bright M dwarfs of \citet{LepineGaidos.2011}. All
stars are selected from the SUPERBLINK catalog of stars with proper
motions $\mu>40$ mas yr$^{-1}$. Stars are identified as probable M
dwarfs based on various color and reduced proper motions cuts; all
selected candidates have, e.g., optical-to-infrared colors
$V-J>2.7$. The low proper motion limit of the SUPERBLINK catalog
excludes nearly all background red giants. The low proper motion
limit however also results in a kinematic bias, whose effects are
discussed in \S2.3 below.

While some astrometric and photometric data have already been compiled
for all the stars, most lack formal spectral classification. Spectral
subtypes have been estimated in \citet{LepineGaidos.2011} only based on a
calibrated relationship between M subtype and $V-J$ color. However,
the $V$ magnitudes of many SUPERBLINK catalog stars are based on
photographic measurements (from POSS-II plates); the resulting $V-J$
colors have relatively low accuracy and are sometimes
unreliable. Besides from affecting spectral type estimates, unreliable
colors can cause contamination of our sample of M dwarf candidates by
bluer G and K dwarfs, which would have otherwise failed the color
cut. These are strong arguments for performing systematic spectroscopic
follow-up observations, to provide reliable spectral typing and filter
out G/K dwarfs (or any remaining M giant contaminants).

A subsample of the brightest of the M dwarfs candidates, with apparent
infrared magnitude J$<9$ was assembled for the first phase of this
survey. We also restricted the sample to stars north of the celestial
equator. This initial list contains a total of 1,564 candidates. All
stars were indiscriminately targeted for follow-up observations,
whether or not they already had well-documented spectra. This would
ensure completeness and uniformity, and allows comparison of our
sample with previous surveys. In particular, our target list
includes M dwarf classification standards from \citet{KHM91} which
provide a solid reference for our spectral classification. The list
includes 557 stars that were previously observed in the PMSU
spectroscopic survey, and 161 that were obseved and classified as part
of the MCN spectroscopic program (including 82 stars observed in both
the PMSU and MCN).

Of the 1,564 M dwarf candidates, we found that 286 had been observed
at the MDM observatory by one of us (SL) prior to November 2008, as
part of a separate spectroscopic follow-up survey of very nearby
(d$<$20pc) stars \citep{Alpert2011}. The remaining targets were
distributed between our observing teams at the MDM Observatory 
(hereafter MDM) and University of Hawaii 2.2-meter Telescope
(hereafter UH), with the MDM team in charge of higher declination
targets ($\delta>30$) and the UH team in charge of the lower
declination range ($0<\delta<30$). In the end we obtained spectra for
all 1,564 stars from the initial target list.

To check for any possible systematic differences arising from using
different telescopes and instruments, we observed 146 stars at both
MDM and UH. We call this subset the ``inter-observatory
subset''. Observations were obtained at different times at the two
observatories. Data were processed in the same manner as the rest of
the sample. 

\begin{figure}
\vspace{-0.4cm}
\hspace{-0.8cm}
\includegraphics[scale=0.47]{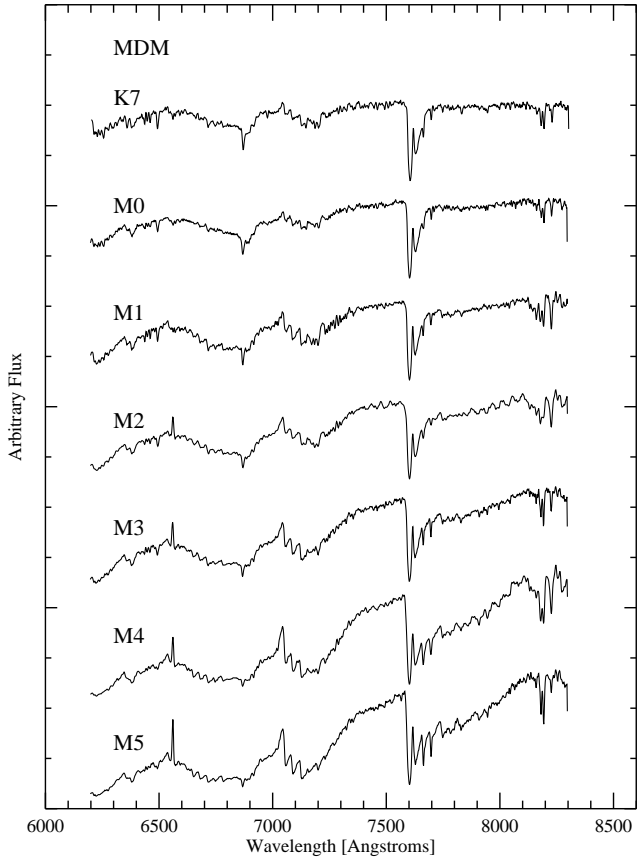}
\caption{Examples of spectra collected at the MDM Obsevatory with
  either the MkIII or CCDS low-resolution spectrographs, and using the
  McGraw-Hill 1.3-meter or Hiltner 2.4-meter telescope. All spectra 
  covered the 6,200\AA-8,700\AA\ wavelength range with a resolution of
  1.8\AA-2.4\AA\ per pixel, and a signal-to-noise ratio
  20$<$S/N$<$30.\label{spec_mdm}}
\end{figure}

\begin{figure}
\vspace{-0.4cm}
\hspace{-0.8cm}
\includegraphics[scale=0.47]{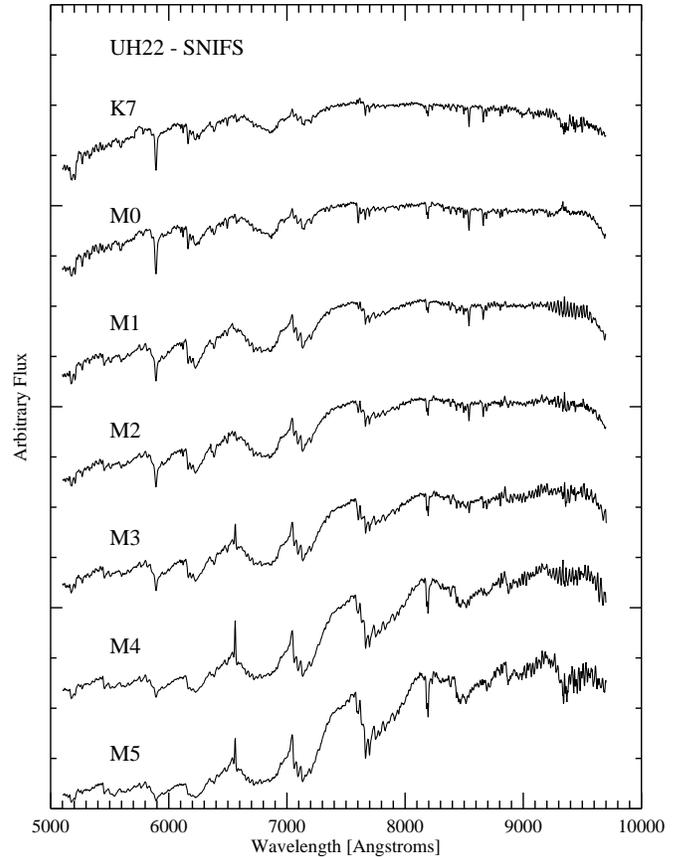}
\caption{Examples of spectra collected at the University of Hawaii
  2.2-meter telescope with the SNIFs spectrograph. Observations
  covered the 5,200\AA-9,800\AA\ wavelength regime with a spectral
  resolution R$\simeq$2000, and a signal-to-noise ratio
  40$<$S/N$<$50.\label{spec_uh}}
\end{figure}


The full list of observed stars is presented in
Table~\ref{table_photo}. We used the standard SUPERBLINK catalog name
as primary designation, however we also include the more widely used
designations (GJ, Gl, and Wo numbers) for the 682 stars listed in the
CNS3 \citep{gj1991}. The CNS3 stars are often well-studied objects,
with abundant data from the literature; a majority of them (580) were
classified as part of the PSMU survey or MCN spectroscopic
program. Another 56 stars on our table which are not in the CNS3 were
however classified as part of the PMSU survey or the MCN program. The
remaining 821 stars are not in the CNS3, and were also not classified
as part of the PMSU survey or MCN progra; these are new
identifications for the most part, and little data existed about them
until now.

Table~\ref{table_photo} lists coordinates and proper motion vectors
for all the stars, along with astrometric parallaxes whenever
available from the literature. The table also lists X-ray source
counts from the ROSAT all-sky point source catalogs
\citep{Voges.etal.1999,Voges.etal.2000}, far-UV ($FUV$) and near-UV
($NUV$) magnitudes from the GALEX fifth data release, optical $V$
magnitude from the SUPERBLINK catalog \citep{LepineShara.2005}, and
infrared $J$, $H$, and $K_s$ magnitudes from 2MASS
\citep{Cutri.etal.2003}. More details on how those data were compiled
can be found in \citet{LepineGaidos.2011}. The optical ($V$ band)
magnitudes listed in Table~\ref{table_photo} come from two sources
with different levels of accuracy and reliability. For 919 stars in
Table~\ref{table_photo}, generally the brightest ones, the $V$
magnitudes come from the Hipparcos and Tycho-2 catalogs. These are
generally more reliable with typical errors smaller than $\pm$0.1
magnitude; those stars are flagged ``T'' in
Table~\ref{table_photo}. The other 645 objects have their $V$
magnitudes estimated from POSS-I and/or POSS-II photographic
magnitudes as prescribed in \citet{Lepine2005}. Photographic
magnitudes of relatively bright stars often suffer from large errors
at the $\sim$0.5 magnitude level or more, in part due to photographic
saturation; those stars are labeled ``P'' in Table~\ref{table_photo}.
 
\begin{deluxetable}{ccccc}
\tabletypesize{\scriptsize}
\tablecolumns{6} 
\tablewidth{0pt} 
\tablecaption{Definition of Spectral Indices.\label{tab:si}}
\tablehead{
\colhead{Index} & 
\colhead{Numerator} &
\colhead{Denominator} &
\colhead{Reference}
}
\startdata 
CaH2   & 6814-6846 & 7042-7046 & \citet{Reid1995} \\
CaH3   & 6960-6990 & 7042-7046 & \citet{Reid1995} \\
TiO5   & 7126-7135 & 7042-7046 & \citet{Reid1995} \\
VO1    & 7430-7470 & 7550-7570 & \citet{Hawley2002} \\
TiO6   & 7550-7570 & 7745-7765 & \citep{Lepine2003} \\
VO2    & 7920-7960 & 8130-8150 & \citep{Lepine2003} 
\enddata     
\end{deluxetable} 

\subsection{Observations}

Spectra were collected at the MDM observatory in a series of 22
observing runs scheduled between June, 2002 and April, 2012. Most of
the spectra were collected at the McGraw-Hill 1.3-meter telescope, but
a number were obtained at the neighboring Hiltner 2.4-meter
telescope. Two different spectrographs were used: the MkIII
spectrograph, and the CCDS spectrograph. Both are facility
instruments which provide low- to medium-resolution spectroscopy in
the optical regime. Their operation at either 1.3-meter or 2.4-meter
telescopes is identical. Data were collected in slit spectroscopy
mode, with an effective slit width of 1.0\arcsec to 1.5\arcsec. The
MkIII spectrograph was used with two different gratings: the 300 l/mm
grating blazed at 8000\AA, providing a spectral resolution
R$\simeq$2000, and the 600 l/mm grating blazed at 5800\AA, which
provides R$\simeq$4000. The two gratings were used with either one of
two thick-chip CCD cameras ({\it Wilbur} and {\it Nellie}) both having
negligible fringing in the red. Internal flats were used to calibrate
the CCD response. Arc lamp spectra of Ne, Ar, and Xe provided
wavelength calibration, and were obtained for every pointing of the
telescope to account for flexure in the system. The spectrophotometric
standard stars Feige 110, Feige 66, and Feige 34 \citep{Oke1991} were
observed on a regular basis to provide spectrophotometric
calibration. Integration times varied depending on seeing, telescope
aperture, and target brightess, but were typically in the 30 seconds
to 300 seconds range. Between 25 and 55 stars were observed on a
typical night. Spectra for a total of 901 bright M dwarf targets were
collected at MDM.

Additional spectra were obtained with the SuperNova Integral Field
Spectrograph \citep[SNIFS; ][]{Lantz2004} on the University of Hawaii
2.2~m telescope on Mauna Kea between February 2009 and November 2012.
SNIFS has separate but overlapping blue (3200-5600\AA) and red
(5200-10000\AA) spectrograph channels, along with an imaging channel,
mounted behind a common shutter. The spectral resolution is
$\sim$1000 in the blue channel, and $\sim$1300 in the red channel; the
spatial resolution of the 225-lenslet array is 0.4\arcsec. SNIFS
operates in a semi-automated mode, acquiring acquisition images to
center the target on the lenslet array, and bias images and
calibration lamp spectra before and after each target spectrum. Both
twilight and dome flats were also obtained every night. Integration
times depended on $J$ magnitude but were 54~s for the faintest ($J=9$)
targets. Up to 75 target spectra were obtained in one night. Spectra
for 655 bright M dwarf targets were collected at UH with SNIFS.

\begin{deluxetable*}{lcrrrrrrrrcrrr}
\tabletypesize{\scriptsize}
\tablecolumns{14} 
\tablewidth{0pt} 
\tablecaption{Survey stars: spectroscopic data \label{table_spectro}}
\tablehead{
\colhead{Star name} & 
\colhead{Observatory} &
\colhead{Julian Date} &
\colhead{CaH2$_{c}$\tablenotemark{a}} &
\colhead{CaH3$_{c}$} &
\colhead{TiO5$_{c}$} &
\colhead{VO1$_{c}$} &
\colhead{TiO6$_{c}$} &
\colhead{VO2$_{c}$} &
\colhead{Sp.Ty.\tablenotemark{b}} &
\colhead{Sp.Ty.} &
\colhead{$\zeta$} &
\colhead{log g\tablenotemark{d}} &
\colhead{$T_{eff}$\tablenotemark{d}} \\
\colhead{} &
\colhead{} &
\colhead{2,450,000+} &
\colhead{} &
\colhead{} &
\colhead{} &
\colhead{} &
\colhead{} &
\colhead{} &
\colhead{index} &
\colhead{adopted} &
\colhead{} &
\colhead{} &
\colhead{K} 
}
\startdata  
PM I00006+1829  &  MDM  & 4791.75 &\nodata&\nodata&\nodata&\nodata&\nodata&\nodata&\nodata&   G/K&\nodata&\nodata&\nodata\\
PM I00012+1358S &  UH22 & 5791.05 & 0.706 & 0.864 & 0.788 & 0.967 & 0.944 & 0.970 &  0.14 &  M0.0&  1.08 &   5.0 &  3790 \\
PM I00033+0441  &  UH22 & 5791.05 & 0.580 & 0.797 & 0.679 & 0.959 & 0.888 & 0.939 &  1.38 &  M1.5&  0.93 &   4.5 &  3510 \\
PM I00051+4547  &  MDM  & 5095.80 & 0.603 & 0.824 & 0.664 & 0.956 & 0.911 & 1.014 &  1.10 &  M1.0&  1.10 &   4.5 &  3560 \\
PM I00051+7406  &  MDM  & 5812.87 &\nodata&\nodata&\nodata&\nodata&\nodata&\nodata&\nodata&   G/K&\nodata&\nodata&\nodata\\
PM I00077+6022  &  MDM  & 5838.82 & 0.364 & 0.620 & 0.392 & 0.905 & 0.630 & 0.798 &  4.60 &  M4.5&  0.90 &   5.0 &  3140 \\
PM I00078+6736  &  MDM  & 5099.83 & 0.534 & 0.789 & 0.623 & 0.905 & 0.806 & 0.929 &  2.03 &  M2.0&  0.96 &   5.0 &  3500 \\
PM I00081+4757  &  MDM  & 5098.84 & 0.410 & 0.700 & 0.420 & 0.871 & 0.684 & 0.814 &  3.80 &  M4.0&  1.01 &   5.0 &  3280 \\
PM I00084+1725  &  UH22 & 5791.06 & 0.671 & 0.842 & 0.785 & 0.970 & 0.947 & 0.970 &  0.34 &  M0.5&  0.91 &   4.5 &  3600 \\
PM I00088+2050  &  MDM  & 4413.72 & 0.372 & 0.646 & 0.356 & 0.893 & 0.602 & 0.793 &  4.58 &  M4.5&  0.99 &   5.0 &  3130 \\
PM I00110+0512  &  UH22 & 5050.03 & 0.653 & 0.839 & 0.706 & 0.962 & 0.893 & 0.925 &  0.86 &  M1.0&  1.16 &   4.5 &  3660 \\
PM I00113+5837  &  MDM  & 5811.90 &\nodata&\nodata&\nodata&\nodata&\nodata&\nodata&\nodata&   G/K&\nodata&\nodata&\nodata\\
PM I00118+2259  &  UH22 & 5050.03 & 0.439 & 0.729 & 0.424 & 0.923 & 0.694 & 0.798 &  3.50 &  M3.5&  1.10 &   4.5 &  3260 \\
PM I00125+2142En&  UH22 & 5791.06 & 0.721 & 0.865 & 0.851 & 0.973 & 0.967 & 0.988 & -0.18 &  M0.0&  0.80 &   4.5 &  3690 \\
PM I00131+7023  &  MDM  & 5812.89 & 0.643 & 0.816 & 0.754 & 0.973 & 0.883 & 1.038 &  0.93 &  M1.0&  0.88 &   4.5 &  3570 
\enddata
\tablenotetext{a}{All spectral indices are corrected for instrumental effects, see Section 3.2.}      
\tablenotetext{b}{Numerical spectral subtype M evaluated from the corrected spectral band indices (not-rounded).}
\tablenotetext{c}{H$\alpha$ spectral index, comparable to equivalent width.}
\tablenotetext{d}{Gravity and effective temperature from PHOENIX model fits.}
\end{deluxetable*} 

Spectroscopic data and results are summarized in
Table~\ref{table_spectro}. SUPERBLINK names are repeated in the first
column and provide a means to match with the entries in
Table~\ref{table_photo}. The second and third columns list the
observatory and Julian date of the observations. The various
spectroscopic measurements whose values are listed in the remaining
columns are described in detail in the sections below.

\subsection{Reduction}

\begin{figure}
\epsscale{1.15}
\plotone{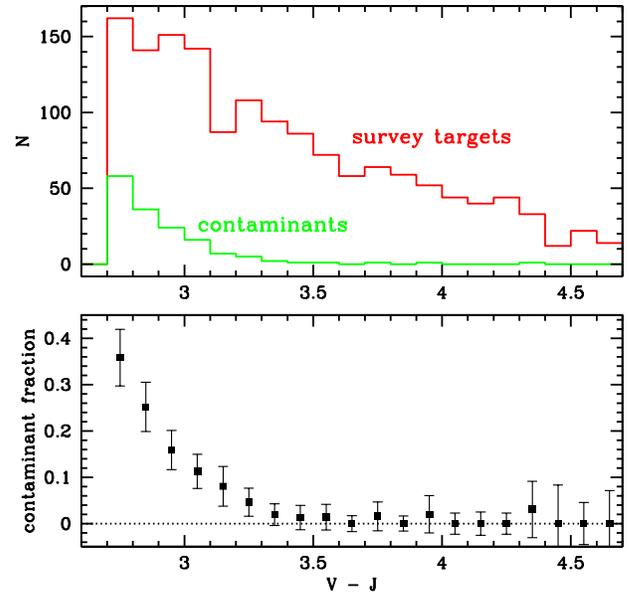}
\caption{Histogram of the $V-J$ color distribution of survey stars
  with spectral morphologies inconsistent with red dwarf of subtype K5
  and later (in green). The distribution of $V-J$ colors from the full
  sample is shown in red. Stars with spectra inconsistent with red
  dwarfs are close to the blue edge of the survey, which indicates
  that they are most probably contaminants of the color-magnitude
  selection, and not the result of mis-acquisition at the
  telescope. \label{contaminant}}
\end{figure}

\subsubsection{MDM data}

Spectral reduction of the MDM spectra was performed using the CCDPROC
and DOSLIT packages in IRAF. Reduction included bias and flatfield
correction, removal of the sky background, aperture extraction, and
wavelength calibration. The spectra were also extinction-corrected and
flux-calibrated based on the measurements obtained from the
spectrophotometric standards. We did not attempt to remove telluric
absorption lines from the spectra. Many spectra were
collected on humid nights or with light cirrus cover, which resulted
in variable telluric features. However, telluric features generally do
not affect standard spectral classification or the measurement of
spectral band indices, since all the spectral indices and primary
classification features avoid regions with telluric absorption. 

A more common problem at the MDM observatory was slit loss
from atmospheric differential refraction. Although this problem could
have been avoided by the use of a wider slit, the concomitant loss of
spectral resolution was deemed more detrimental to our science
goals. Instead, stars were observed as close to the meridian as
observational constraints allowed. In some cases, however, stars were
observed up to $\pm$2 hours from the meridian, resulting in noticeable
slit losses. Fluctuations in the seeing, which often exceeded the slit
width, played a role as well. As a result, the spectrophotometric
calibration was not always reliable, since the standards were only
observed once per night to maximize survey efficiency. Flux
recalibration was
therefore performed using the following procedure. Spectral indices
were measured and the spectra were classified using the classification
method outlined below (\S 3.1). The spectra were then compared to
classification templates assembled from M dwarf spectra from the Sloan
Digital Sky Survey (SDSS) spectroscopic database. Each spectrum
was divided by the classification template of the same alleged
spectral subtype. In many cases, the quotients yielded a flat
function, indicating that the spectrophotometric calibration was
acceptable. Other quotients yielded residuals consistent with first or
second order polynomials spanning the entire wavelength range, and
indicating problems in the spectrophotometric calibration. A
second-order polynomial was fit through the quotient spectra, and the
original spectrum was normalized by this fit, correcting for
calibration errors. Spectral indices were then measured again, and the
spectra reclassified; this yielded changes by 0.5 to 1.0 subtypes for
$\approx20\%$ of the stars (no changes for the rest). The
re-normalization was then performed again using the revised spectral
subtypes, and the procedure repeated until convergence for all stars.

Finally, all the spectra were wavelength-shifted to the rest frames
of their emitting stars. This was done by cross-correlating each
spectrum with the SDSS template of the corresponding spectral
subtype. Spectral indices were again re-measured, and the stars
re-classified. Any change in the spectral subtype prompted a repeat of
the flux-recalibration procedure outlined above, and the
cross-correlation procedure was repeated using the revised spectral
template until converegence.

A sequence of MDM spectra is displayed in Figure~\ref{spec_mdm},
which illustrates the wavelength regime and typical quality. Note the
telluric absorption features near 6,850\AA, 7,600\AA, and
8,200\AA. Signal-to-noise ratio is generally in the 30$<$S/N$<$50
range near 7500\AA.

\begin{figure*}
\vspace{-6.5cm}
\hspace{0.3cm}
\includegraphics[scale=0.87]{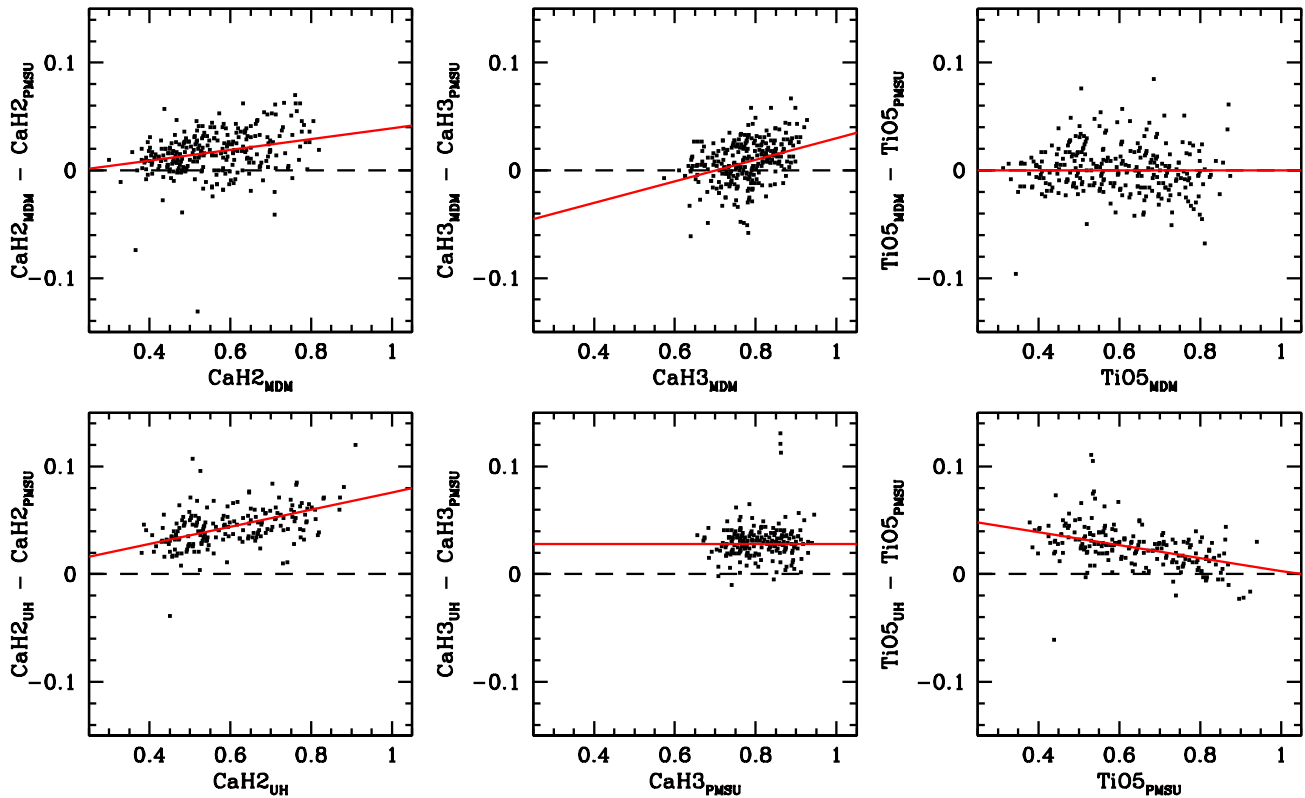}
\caption{Comparison between our spectral band index measurements for a
  subset of 484 stars observed at MDM and UH, and the band index
  measurements for the same stars as reported in the Palomar-MSU
  spectroscopic survey of \citet{Reid1995}. Small but systematic offsets
  are observed, which are explained by differences in spectral
  resolution and spectrophotometric calibration between the different
  observatories. These offsets demonstrate that the spectral indices
  are instrument-dependent, but that the measurements can be corrected
  by observing large subsets of stars at the different
  obervatories. The red segments show the fits to the offsets, which
  are used to calibrate corrections for each observatory, here using
  the Palomar-MSU measurements as the standard.\label{pmscomp}}
\end{figure*}

\subsubsection{SNIFS data}

SNIFS data processing was performed with the SNIFS data reduction
pipeline, which is described in detail in \citet{Bacon:2001} and
\citet{Aldering:2006}. After standard CCD preprocessing (dark, bias,
and flat-field corrections), data were assembled into red and blue 3D
data cubes. The data cubes were cleaned for cosmic rays and bad
pixels, wavelength-calibrated using arc lamp exposures acquired
immediately after the science exposures, and spectrospatially
flat-fielded, using continuum lamp exposures obtained during the same
night. The data cubes were then sky-subtracted, and the 1D spectra were
extracted using a semi-analytic PSF model. We applied corrections to
the 1D spectra for instrument response, airmass, and telluric lines
based on observations of the Feige 66, Feige 110, BD+284211, or
BD+174708 standard stars \citep{Oke1991}. Because the SNIFS spectra
are from an integral field spectrograph, operating without a slit,
their spectrophotometry is significantly more reliable than the
slit-spectra obtained at MDM. In fact, it is possible to perform
synthetic photometry by convolving with the proper filter response.

As with the MDM data, SNIFS spectra were shifted to the rest frames of
their emitting stars by cross-correlation to SDSS templates
\citep{Bochanski2007} of the corresponding spectral subtype. Spectral
indices were re-measured and the stars re-classified. This process was
repeated if there was a change in the spectral subtype determination.

A sequence of UH SNIFS spectra are displayed in Figure~\ref{spec_uh},
which show the wavelength range and typical data
quality. Signal-to-noise ratio is generally S/N$\approx$100 near
7500\AA.

\section{Spectral classification}

\subsection{Visual identification of contaminants}

We first examined all the spectra by eye to confirm the presence of
morphological features consistent with red dwarfs of spectral subtype
K5 and later. The main diagnostic was the detection of broad CaH and
TiO moleculars band near 7000\AA. Of the 1564 stars observed, 1408 were
found to have clear evidence of CaH and TiO. The remaining 156 stars
did not show those molecular features clearly, within the noise
limit, and were therefore identified as probable contaminants in the
target selection algorithm.

Most of these contaminants appeared to be early to mid-type K dwarfs,
with a few looking like G dwarfs affected by interstellar reddening. A
number of stars also displayed carbon features consistent with low gravity
objects, most probably K giants. We suspect that many of the G and
K dwarfs have inaccurate V-band magnitudes, making them appear redder
than they really are, which would explain their inclusion in our
color-selected sample. Interstellar reddening would also explain the
inclusion of more distant G dwarfs in our sample, due to their redder
colors. An alternate explanation, however, might be that the targets
were mis-acquired in the course of the survey, and that the spectra
represent random field stars. Indeed the very large proper motion of
the sources sometimes makes them difficult to identify at the
telescope, as they often have moved significantly from their positions
on finder charts. Stars in crowded field are particularly susceptible
to this effect. To verify this hypothesis, we compared the $V-J$ color
distribution of the contaminants to the distribution of the full
survey sample (Figure~\ref{contaminant}, top panel); the fraction of
contaminant stars in each color bin is also shown (bottom panel). The
two distributions are significantly different, with the contaminants
being dominated by relatively blue stars, and their fraction quickly
drops as $V-J$ increases. We can only conclude that the contaminants
are not mis-acquired stars, otherwise one would expect the two
distributions to be statistically equivalent. Rather, the majority of
the contaminants must have been properly acquired and are simply
moderately red FGK stars that slipped into the sample in the
photometric/proper motion selection, as suggested first.

\begin{figure*}
\vspace{-6.5cm}
\hspace{0.3cm}
\includegraphics[scale=0.87]{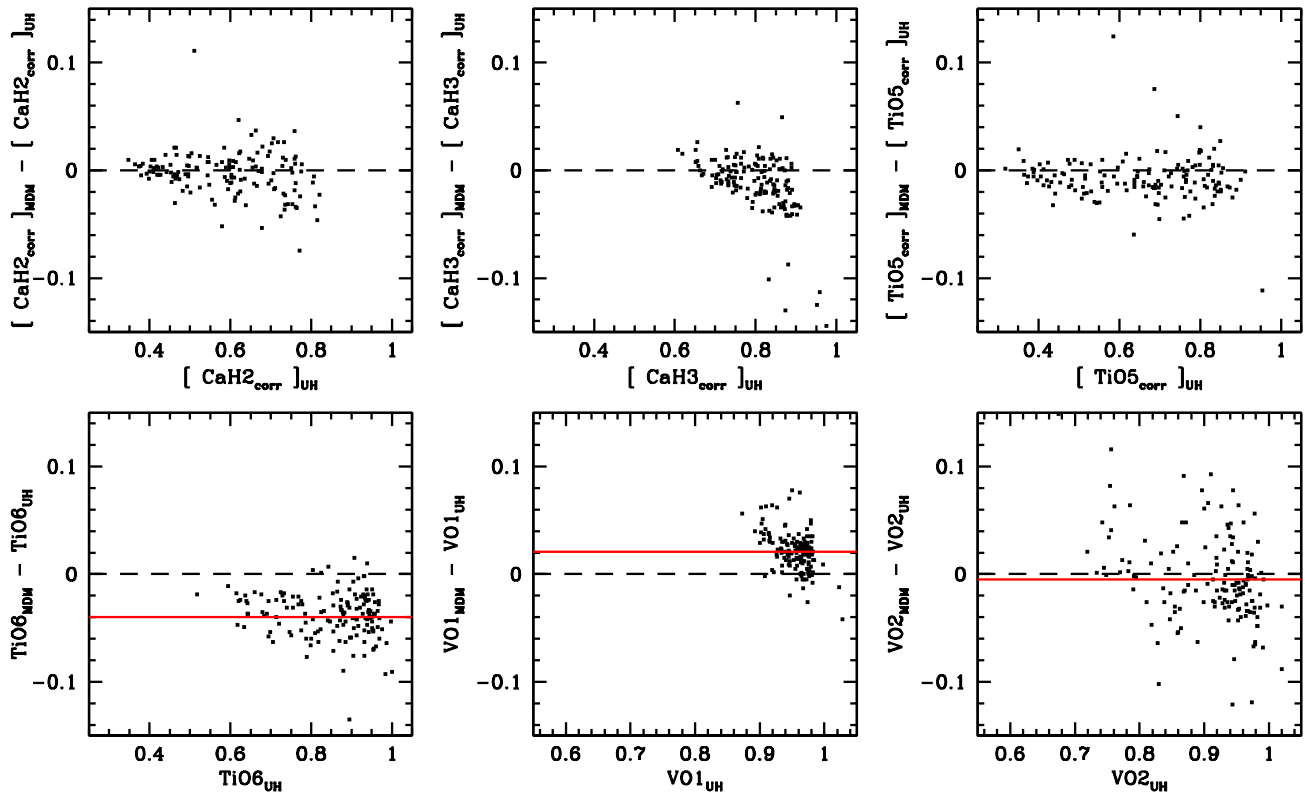}
\caption{Comparison between the {\em corrected} spectral band index
  measurements from MDM and the band index measurements of the same
  stars observed from UH. Observatory-specific corrections to the
  CaH2, CaH3, and TiO5 indices (see Figure~\ref{pmscomp}) bring the
  MDM and UH values in close agreement. Small offsets are however
  observed for the TiO6, VO1, and VO2 indices, which we could not
  calibrate against Palomar-MSU survey stars. The horizontal red lines
  show the adopted offsets which are used to correct the MDM values to
  bring them in line with the UH ones. Offsets are again believed to
  be due to inter-observatory differences in the spectral resolution
  and spectrophotometric calibration.\label{idx_comp}}
\end{figure*}

We find an overall contamination rate of $\simeq$10\% in our survey,
although most of the contamination occurs among stars with relatively
blue colors. The contamination rate is $\simeq$26\% for red dwarf
candidates in the $2.7<V-J<3.0$ color range, but this rate drops to
$\simeq$8\% for candidates with $3.0<V-J<3.3$, and becomes negligible
($<1\%$) in the redder $$V-J>3.3$$ candidates. The 156 stars
identified as contaminants are included in Table~\ref{table_photo} and
Table~\ref{table_spectro} for completeness and future
verification. Spectroscopic measurements for these stars, such as band
indices, subtypes, and effective temperatures, are however left blank.

\subsection{Classification by spectral band indices}

\subsubsection{M dwarf classification from molecular bandstrengths}

The spectra of M dwarfs are dominated by molecular bands from metal
oxides (mainly TiO, VO), metal hydrides (CaH, CrH, FeH), and metal
hydroxides (CaOH). The most prominent of these in the optical-red
wavelength range (5000\AA-9,000\AA) are the bands from titanium oxide
(TiO) and calcium hydride (CaH). The resulting opacities from those
broad moleciular bands significantly affect the broadband colors and
spectral energy distribution of M dwarfs
\citep{Jones1968,Allard2000,Kraw2000}. Early atmospheric models of M
dwarfs showed that the strength of the TiO and CaH bands depends on
effective temperature, but also on surface gravity and metal
abundances \citep{Mould.1976}. 

Molecular bands have historically been the defining diagnostic and
classification features of M dwarfs. For stars that have settled on
the main sequence, one can assume that the surface gravity is entirely
constrained by the mass and chemical composition. Leaving only the
effective temperature and chemical abundances as general parameters in
the classification and/or spectroscopic modeling. For local disk stars
of solar metallicity, a classification system representing an
effective temperature sequence can thus be established based on
molecular bandstrentghs.

The detection and measurement of TiO and CaH molecular bands thus
forms the basis for the M dwarf classification system
\citep{JoyAbt1974}. Molecular bands become detectable starting at
spectral subtype K5 and K7, the latest subtypes for K dwarfs (there
are no K6, K8, or K9 subtypes). The increasing strength of the
molecular bands then defines a sequence running from M0 to M9. The
strength of both TiO and CaH molecular bands reach a maximum around
$T_{eff}\simeq$2700K. There is a turnaround in the correlation below
this point, and molecular bands become progressively weaker at lower
temperatures until they vanish \citep{CruzReid.2002}. The reversal and
weakening is thought to be due to the condensation of molecules into
dust, and their settling below the photosphere \citep{JonesTsuji1997}.

\begin{deluxetable*}{lcccccc}
\tabletypesize{\scriptsize}
\tablecolumns{7} 
\tablewidth{0pt} 
\tablecaption{Coefficients of the spectral index corrections, by
  observatory.\label{tab:sic}}
\tablehead{
\colhead{{\it IDX}} & 
\colhead{$A_{IDX:{\rm PMSU}}$} &
\colhead{$B_{IDX:{\rm PMSU}}$} &
\colhead{$A_{IDX:{\rm MDM}}$} &
\colhead{$B_{IDX:{\rm MDM}}$} &
\colhead{$A_{IDX:{\rm UH}}$} &
\colhead{$B_{IDX:{\rm UH}}$}
}
\startdata 
CaH2   &   1.00 &   0.00 & 0.95 &  0.011 & 0.92 &  0.004 \\ 
CaH3   &   1.00 &   0.00 & 0.90 &  0.070 & 1.00 & -0.028 \\
TiO5   &   1.00 &   0.00 & 1.00 &  0.000 & 1.06 & -0.063 \\
VO1    & \nodata& \nodata& 1.00 &  0.040 & 1.00 &  0.000 \\
TiO6   & \nodata& \nodata& 1.00 & -0.021 & 1.00 &  0.000 \\
VO2    & \nodata& \nodata& 1.00 &  0.005 & 1.00 &  0.000
\enddata     
\end{deluxetable*} 

Sequences of classification standards were compiled in \citep{KHM91},
which identified the main molecular bands in the yellow-red spectral
regime, where TiO and CaH bandheads are most prominent. To better
quantify the classification system, a number of ``band indices'' were
defined by \citet{Reid1995}, which measure the ratio between on-band and
off-band flux, for various molecular bandheads. Calibration of these
band indices against classification standards provide a means to
objectively assign subtypes based on spectroscopic
measurements. Originally, these band indices measured the strengths of
various features near $7,000\AA$, where the most prominent CaH and TiO
features are found in early-type M dwarfs. These bands, however,
become saturated in late-type M dwarfs, which makes their use
problematic in later dwarfs. There is a VO band near $7,000$\AA,
located just between the main CaH and TiO features, which becomes
prominent only at later subtypes; and band index measuring this
feature was introduced as a primary diagnosis for the so-called
``ultra-cool'' M dwarfs, and provided a classification scale for
subtypes M7-M9 \citep{KHS95}. Additional band indices associated with
TiO and VO bands in the 7500\AA-9000\AA\ range, where molecular
absorption develops at later subtypes, were introduced as secondary
classification features \citep{Lepine2003}. These form the basis of
current classification methods based on optical-red spectroscopy.

Nearby low-mass stars associated with the local halo population have
long been known to show peculiar banstrength ratios, and in particular
to have weak TiO compared to CaH
\citep{Mould.1976,Mould_McElroy.1978}. A system to identify and
classify the metal-poor M subdwarfs based on the strength and ratio of
CaH and TiO was introduced by \citet{Gizis1997}, and expanded by
\citet{Lepine2003b} and \citet{Lepine2007}, as the spectroscopic
census of M subdwarfs grew larger. In this system, the CaH
bandstrengths are used as a proxy of $T_eff$ and determine spectral
subtypes, while the TiO/CaH band ratio is used to evaluate
metallicity. For that purpose, the $\zeta_{TiO/CaH}$ parameter, which
is a function of the TiO/CaH band ratio, was introduced by
\citet{Lepine2007} as a possible proxy for metallicity, and a
tentative calibration with [Fe/H] was presented by \citet{Woolf2009}.
One of the main issues in the current M dwarf classification scheme,
is that both TiO and CaH bandstrengths are used to determine the
spectral subtype, whereas TiO is now believed to be quite sensitive to
metallicity. This means, e.g., that moderately metal-rich M dwarfs may
be assigned later subtypes than Solar-metallicity ones. In addition,
the classification of young field M dwarfs may be affected by their
lower surface gravities, which also tend to increase TiO
bandstrengths and would thus yield to marginally later subtype
assignments compared with older stars of the same $T_{eff}$. These
caveats must be considered when one uses M dwarf spectral subtypes as
a proxy for surface temperature.

\begin{figure*}
\vspace{-6.5cm}
\hspace{0.3cm}
\includegraphics[scale=0.87]{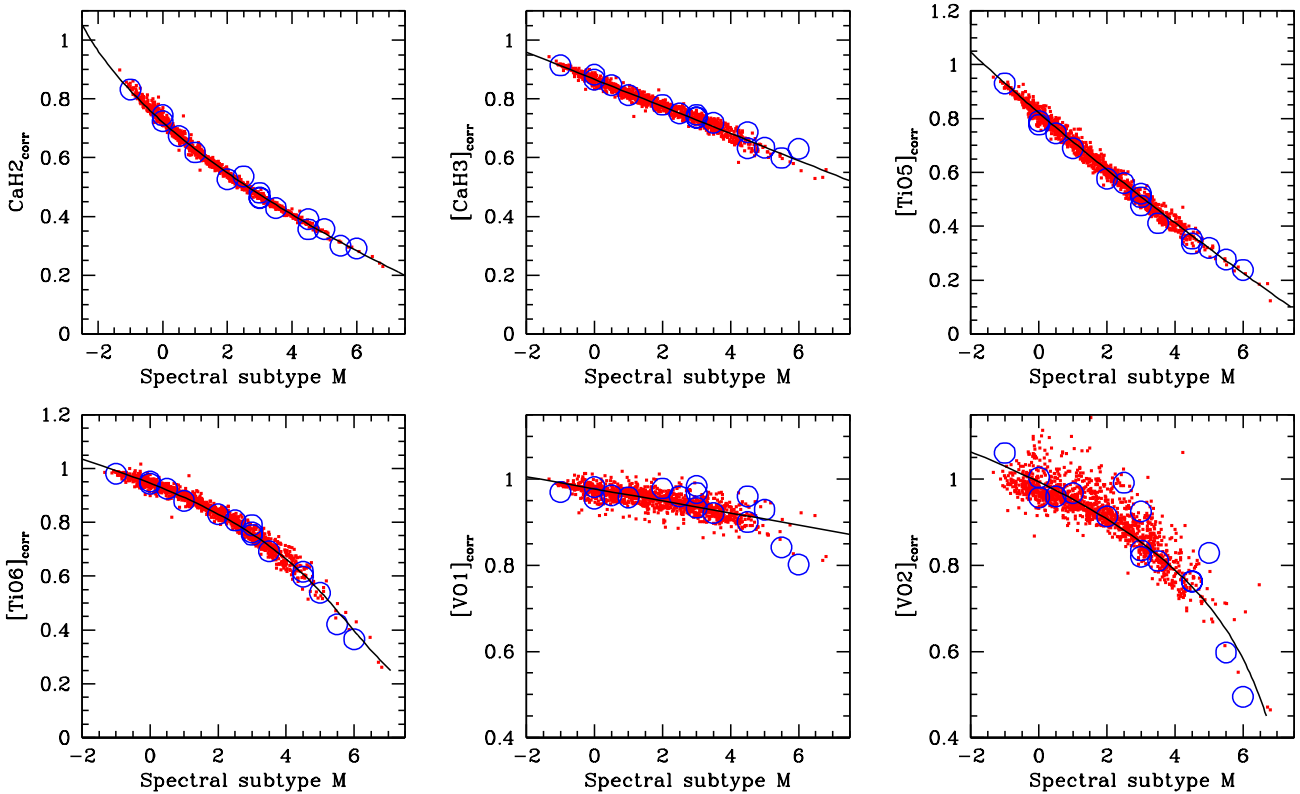}
\caption{Variation in the corrected spectral indices as a function 
  of adopted spectral subtypes. The thin black lines show the
  adopted, revised calibrations for spectral
  classification. Blue circles represent the subset of classification
  standards from \citet{KHM91} which we observed; the X-axis values of
  those data points are the formal spectral subtypes adopted in
  \citet{KHM91}, while the Y-axis values are the spectral index
  measurements from our survey. For our own spectral typing scheme, we
  calculate the average of the subtype values for the ${\rm
    CaH2}_{c}$, ${\rm CaH3}_{c}$, ${\rm TiO5}_{c}$, and ${\rm
    TiO6}_{c}$ indices given by the adopted relationships. The two VO
  indices have relatively weak leverage on early type stars due to the
  shallow slope in the relationships, and are not used for assigning
  spectral subtypes of our (mostly early-type) survey
  stars.\label{idx_spty}}
\end{figure*}

\subsubsection{Definition and measurement of band indices}

The strength of the TiO, CaH, and VO molecular bands are measured
using spectral band indices. These spectral indices measure the ratio
between the flux in a section of the spectrum affected by molecular
opacity to the flux in a neighboring section of the spectrum minimally
affected by molecular opacity. The latter section defines a
pseudo-continuum of sorts, although M dwarf spectra do not have a
continuum in the classical sense, because their spectral energy
distribution strongly deviates from that of a blackbody, and is
essentially shaped by atomic and molecular line opacities.

We settle on a set of six spectral band indices: CaH2, CaH3, TiO5,
TiO6, TiO7, VO1, and VO2. These band indixes, which we previously used
in \citet{Lepine2003} to classify spectra collected at MDM, measure
the strength of the most prominent bands of CaH, TiO, and VO in the
$6000{\rm   \AA}<\lambda<8500{\rm \AA}$ regime. The spectral indices
and are listed in Table~\ref{tab:si} along with their definition. The
CaH2, CaH3, and TiO5 indices are the same as those used in the
Palomar-MSU survey, and were first defined in \citep{Reid1995}. The TiO6,
VO1, and VO2 indices were introduced by \citet{Lepine2003} to better
classify late-type M dwarfs, whose CaH2 and TiO5 indices become
saturated at cooler tempratuers and are not as effective for accurate
spectral classification of late-type M dwarfs. Each spectral band
index is calculated as the ratio of the flux in the spectral region of
interest (numerator) to the flux in the reference region
(denominator), i.e.: 
\begin{equation} IDX = \frac{\int_{num} I(\lambda) d\lambda}{\int_{denom}
  I(\lambda) d\lambda} 
\end{equation}
Because the wavelength range for some indices is relatively narrow
(especially the denominator for CaH2, CaH3, and TiO5) it is important
that the spectra in which they are measured have their wavelengths
calibrated in the rest frame of the star, which is why special care
was made to correct all spectra for any significant redshift/blueshift
(see above). 

Because the measured molecular bandheads are relatively sharp, and
because the spectral indices measuring them are defined over
relatively narrow spectral ranges, the index values are potentially
dependent on the spectroscopic resolution, and may thus depend
on the specific instrumental setup used for the observations. In
addition, the index values may be affected by systematic errors in the
spectrophotometric flux calibration, which can also be dependent on
the instrument and/or observatory where the measurements were
made. One way to verify these effects is to compare spectral index
measurements of the same stars obtained at different
observatories. Because of the significant overlap with the Palomar-MSU
survey, we can use those stars as reference sample, and recalibrate
the spectral indices so that they are consistent to those reported in
\citet{Reid1995}. 

Our census have 557 stars in common with the PMSU spectroscopic
survey; these stars are all identified with a flag (``P'') in the last
column of Table~\ref{table_spectro}. We identify 206 stars from the
list observed at MDM and 281 stars from the list observed at UH which
have spectral index measurements
reported in the PMSU survey. The differences between our measured
CaH2, CaH3, and TiO5 and those reported in the PSMU catalog are
plotted in Figure~\ref{pmscomp}. Trends and offsets confirm the
existence of systematic errors, possibly due to differences in
resolution and flux calibration. To verify the spectroscopic
resolution hypothesis, we convolved the MDM spectra with a box kernel
5-pixel wide; we found that indeed the MDM indices for the smoothed
spectra had their offsets reduced by 0.01-0.02 units, bringing them
more in line with the PMSU indices. We also observe that the MDM
measurements tend to have a larger scatter than the UH ones; we
suggest that this may be due to spectrophometric calibration issues
with some of the MDM spectra, as discussed in \S2.3.1.

To achieve consistency in the measurements obtained at different
observatories, we adopt the values from the PMSU survey as a standard
of reference, and calculate corrections to the measurements from MDM and
UH by fitting linear relationships to the residuals. A corrections to a
spectral band index is thus applied following the general function:
\begin{equation}
IDX_{c} = A_{IDX:OBS}\ IDX_{OBS} + B_{IDX:OBS}  \\ 
\end{equation}
where $IDX_{OBS}$ represents the measured value of an index at the
observatory $OBS$, and ($A_{IDX:OBS}$,$B_{IDX:OBS}$) are the
coefficients of the transformation from the observed value to the
corrected one ($IDX_{c}$). Hence the corrected values of the indices
CaH2, CaH3 and TiO5 for measurements done at MDM are defined as:
\begin{eqnarray}
{\rm CaH2_{c} = A_{\rm CaH2:MDM} CaH2_{\rm MDM} + B_{\rm CaH2:MDM} } \\ 
{\rm CaH3_{c} = A_{\rm CaH3:MDM} CaH3_{\rm MDM} + B_{\rm CaH3:MDM} } \\ 
{\rm TiO5_{c} = A_{\rm TiO5:MDM} TiO5_{\rm MDM} + B_{\rm TiO5:MDM} }  
\end{eqnarray}
The measurements from the PSMU survey are used as standards for these
three indices, and we thus have by definition: $A_{\rm
  CaH2:PMSU}=A_{\rm CaH3:PMSU}=A_{\rm TiO5:PMSU}=1.0$, $B_{\rm
  CaH2:PMSU}=B_{CaH3:PMSU}=B_{\rm TiO5:PMSU}=0.0$.
For OBS=MDM and OBS=UH, the adopted correction coefficients are listed
in Table~\ref{tab:sic}. The corresponding linear relationships are
shown as red segments in Figure~\ref{pmscomp}. 

To verify the consistency of the corrected spectral band index values,
we compare the corrected values for the stars observed at both MDM and
UH (the inter-observatory subset). The differences are shown in
Figure~\ref{idx_comp}. We find the corrected values CaH2$_{c}$,
CaH3$_{c}$, and TiO5$_{c}$ to be in good agreement, with no
significant offsets beyond what is expected from measurement
errors. The corrected values of all three spectral indices are listed
in Table~\ref{table_spectro}.

The TiO6, VO1, and VO2 spectral index mesurements from MDM and UH are
also compared in Figure~\ref{idx_comp}. Those were not measured in the
PMSU survey, since the indices were introduced later
\citep{Lepine2003}, and thus are displayed here without any
correction. Small but significant offsets between the MDM and UH
values again suggest systematic errors due to differences in spectral
resolution and flux calibration. This time we adopt the UH
measurements as fiducials, and determine corrections to be applied to
the MDM data. The corrections are listed in Table~\ref{tab:sic} and
the corresponding linear relationships are displayed in
Figure~\ref{idx_comp} as red segments. The corrected values of the
three spectral indices are also listed in Table~\ref{table_spectro}.

The scatter between the MDM and UH values, after correction, as well
as the scatter between the UH/MDM and PMSU values, provide an
estimate of the measurment accuracy for these spectral
indices. Excluding a few outliers, the mean scatter is $\approx$0.02
units (1$\sigma$) for the CaH2, CaH3, TiO5, TiO6, and VO1 indices, and
$\approx$0.04 units for VO2. This assumes that M stars do not show any
significant changes in their spectral morphology over time, and that
the spectral indices should thus not be variable.

\subsubsection{Spectral subtype assignments for K/M dwarfs}

Because of the correlation between spectral subtype and the depth of
the molecular bands, it is possible to use the values of the spectral
band indices to estimate spectral subtypes. This only requires a
calibration of the relationship between spectral index values and the
spectral subtypes, in a set of stars which were classified by other
means, e.g., classification standards. The system adopted in this
paper uses the spectral indices listed in Table~\ref{tab:si}, and
follows the methodology outlined in \citep{Gizis1997} and
\citep{Lepine2003}. Relationships are calibrated for each spectral
index, and spectral subtypes are calculated from the mean values
obtained from all relevant/available spectral indices. The mean values
are then be rounded to the nearest half integer, to provide formal
subtyping with half-integer resolution. The system is extended to
late-K dwarfs as well: an ``M subtype'' with a value $<0.0$ signifies
that star is a late-K dwarf: the star is classified as K7 for an index
value $\approx-1.0$ and as K5 for an index value $\approx-2.0$ (note:
there is no K6 subtype for dwarf stars, and K7 is the subtype
immediately preceding M0).

The original spectral-index classification method for M
dwarfs/subdwarfs is based on a relationship between subtype and with
the CaH2 index, which measures one of the most prominent band at all
spectral subtypes, and notably displays the deepest bandhead in
metal-poor M subdwarfs \citet{Gizis1997,Lepine2003}. The original
relationship is: $\left[ SpTy \right]_{CaH2} = 10.71 - 20.63
\ {\rm CaH2} + 7.91 \ \left( {\rm CaH2} \right)^2$. To verify this
relationship, we estimated spectral subtypes from our corrected
indices CaH2$_{c}$ for 16 spectroscopic calibration standards from
\citet{KHM91}, which were observed as part of our survey, and span a
range of spectral subtypes from K7.0 to M6.0. We found small but
significant differences in our estimated spectral subtypes and the
values formally assigned by \citet{KHM91}; subtypes estimated from
the \citet{Gizis1997} relationship tend to systematically
underestimate the standard subtypes by $\approx0.5$ units for stars
later than M3. To improve on the index classification method, we
performed a $\chi^2$ polynomial fit to recalibrate the relationship,
obtaining:
\begin{equation}
\left[ \rm SpTy \right]_{\rm CaH2} = 11.50 - 21.71 \ {\rm CaH2}_{c} + 7.99
  \ \left( {\rm CaH2} \right)^2
\end{equation}
which does correct for the observed offsets at later types. Using this
this relatioship as a starting point, and guided by the formal
spectral subtype from the classification standards, we performed
additional $\chi^2$ polynomial fits to calibrate an index-subtype
relationships for CaH3:
\begin{equation}
\left[ \rm SpTy \right]_{\rm CaH3} = 18.80 - 21.68 \ {\rm CaH3}_{c}
\end{equation}
Where the corrected values of the spectral bands indices (see Eqs.2-5)
are used. The relationships are slightly different from those quoted in
\citet{Gizis1997} and \citet{Lepine2003} but are internally
consistent to each other, whereas an application of the older
relationships to our corrected band index measurements would yield
internal inconsistencies, with subtype difference up to 1 spectral
subtype between the relationships.

The ratio of oxides (TiO, VO) to hydrides (CaH,CrH,FeH) in M dwarfs is
known to vary significantly with metallicity
\citep{Gizis1997,Lepine2007}. In the metal-poor M subdwarfs, it is the
oxides bands that appear to be weaker, while hydride bands remain
relatively strong (in the most metal-poor ultrasubdwarfs, or usdM,
the TiO bands are almost undetectable). Therefore it makes sense to
rely more on the CaH band as the primary subtype/temperature
calibrator. The same $[\rm SpTy]_{\rm CaH2}$ and $[\rm SpTy]_{\rm
  CaH3}$ relationships should be used to determine spectral subtypes
at all metallicity classes (i.e. in M subdwarfs as well as in M
dwarfs).

Because the TiO and VO bands are also strong in the metal-rich M
dwarfs, it is still useful to include these bands as secondary
indicators, to refine the spectral classification. In the late-type M
dwarfs, in fact, the CaH bandheads are saturating, and one has to rely
on the TiO and VO bands. In fact, the VO bands were originally used to
diagnose and calibrate ultracool M dwarfs of subtypes M7-M9 \citep{KHS95}.
The main caveat in using the oxide bands for spectral classification
is that this can potentially introduce a metallicity dependence on the
estimated spectral subtype, with more metal-rich stars being assigned
later subtypes than what they would have based on the strength of
their CaH bands alone. In any case, because our sample appears to be
dominated by near-solar metallicity stars, we calibrate additional
relationships between subtype and the TiO5 and TiO6 bands indices. We
first recalculate the subtypes by averaging the values of $\left[ SpTy
  \right]_{CaH2}$ and $\left[ SpTy \right]_{CaH2}$, and perform a
$\chi^2$ fit of the TiO5 and TiO6 indices to the mean subtypews
calculated from CaH2 and CaH3, finding:
\begin{equation}
\left[ \rm SpTy \right]_{\rm TiO5} = 7.83 - 9.55 \ {\rm TiO5}_{c}
\end{equation}
\begin{displaymath}
\left[ \rm SpTy \right]_{\rm TiO6} = 9.92 - 15.68 \ {\rm TiO6}_{c}
+21.23 \ \left( {\rm TiO6}_{c} \right)^2 
\end{displaymath}
\begin{equation}
- 16.65 \ \left( {\rm TiO6}_{c} \right)^3
\end{equation}
where again the corrected band indices are used. The relatively sharp
non-linear deviation in the TiO6 distribution around M3 forces the use
of a third order polynomial in the fit. 

We also determine the relationships for the VO1 and VO2
band indices. This this after recalculating the subtypes from the
average of $\left[ \rm SpTy \right]_{\rm CaH2}$ and $\left[ \rm SpTy
  \right]_{\rm   CaH2}$, $\left[ \rm SpTy \right]_{\rm TiO5}$, and
$\left[ \rm SpTy   \right]_{\rm TiO6}$, a $\chi^2$ fit again yields: 
\begin{equation}
  \left[ \rm SpTy \right]_{\rm VO1} = 69.8 - 71.4 \ {\rm
  \left[\rm VO1\right]_{c}} 
\end{equation}
\begin{displaymath}
\left[ \rm SpTy \right]_{\rm VO2} = 9.56 - 12.47 \ {\rm
  \left[\rm VO2\right]_{c}} 
 + 22.33 \ \left( {\rm \left[\rm VO2\right]_{c}} \right)^2
\end{displaymath}
\begin{equation}
 - 19.59 \ \left( {\rm \left[\rm VO2\right]_{c}} \right)^3.
\end{equation}
The VO indices however make relatively poor estimators of spectral
subtypes for our sample, mainly because the shallow slope at earlier
subtypes provides little leverage. The VO2 index also shows
unexpectedly large scatter in the MDM spectra, including in the
classification standard stars, which we suspect is due the fact that
the index is defined very close to the red edge of the MDM spectral
range and is thus more subject to statistical noise and flux
calibration errors. We therefore do not include $\left[ SpTy
  \right]_{VO12}$ and $\left[ SpTy \right]_{VO2}$ in the final
determination of the spectral subtypes.

\begin{deluxetable}{cc}
\tabletypesize{\scriptsize}
\tablecolumns{2} 
\tablewidth{160pt} 
\tablecaption{Distribution by spectral subtype of the 1564 survey
  stars\label{table_st}}
\tablehead{
\colhead{Spectral subtype} & 
\colhead{N} 
}
\startdata 
G/K\tablenotemark{a} &  160 \\
K7.0   &  27 \\
K7.5   & 101 \\
M0.0   & 177 \\
M0.5   & 160 \\
M1.0   & 152 \\
M1.5   & 147 \\
M2.0   & 141 \\
M2.5   & 119 \\
M3.0   & 125 \\
M3.5   & 125 \\
M4.0   &  72 \\
M4.5   &  36 \\
M5.0   &  13 \\
M5.5   &   4 \\
M6.0   &   2 \\
M6.5   &   2 \\
M7.0   &   1
\enddata     
\tablenotetext{a}{Stars identified as earlier than K7.0 and/or with no
  detected molecular bands.}
\end{deluxetable} 

Fig.~\ref{idx_spty} plots all the corrected spectral band indices as a
function of the adopted spectral subtype. The relatively
small scatter ($\approx0.02$) in the distribution of $\left[\rm
  CaH2\right]_{c}$, $\left[\rm CaH3\right]_{c}$, $\left[\rm
  TiO5\right]_{c}$ and $\left[\rm TiO6\right]_{c}$ demonstrate
the internal consistency of the spectral type calibration for the four
indices. All four relationships have an average slope $\approx10$,
which means that since those indices have an estimated measurement
accuracy of $\approx0.02$ units, the spectral subtypes calculated by
combining the four indices should be accurate to about $\pm0.10$
subtype assuming that the measurement errors in the four indices are
uncorrelated. While this would make it possible to classify the stars
to within a tenth of a subtype, we prefer to follow the general
convention and assign spectral subtypes to the nearest half
integer. 

To verify the consistency of the spectral classification, we
compare the spectral types evaluated independently for the list of 141
stars observed at both MDM and UH. We find that 82\% of the stars end
up with the same spectral type assigments from both observatories,
i.e. they have spectra assigned to the same half-subtype. All the
other stars have classifications within 0.5 subtypes. This is
statistically consistent with a $1\sigma$ error of $\pm$0.18 on the
spectral type determination, slightly larger than the assumed $0.1$
subtype precision estimated above. This suggests a $3\sigma$ error of
about $\pm$0.5, which justifies the more conservative use of
half-subtypes as the smallest unit for our classification.

The resulting classifications based on the CaH and TiO band index
measurements are listed in Table~\ref{table_spectro}. The numerical
spectral subtype measured from the average of the band indices is
listed to 2 decimal figures. These values are rounded to the nearest
half integers to provide our more formal spectral classifications to a
half-subtype precision. The non-rounded values are however useful for
comparison with other physical parameters as they provide a continuous
range of fractional values; these fractional subtype values are used
in the analysis throughout the paper

\begin{figure}
\vspace{-2cm}
\hspace{-0.1cm}
\includegraphics[scale=0.42]{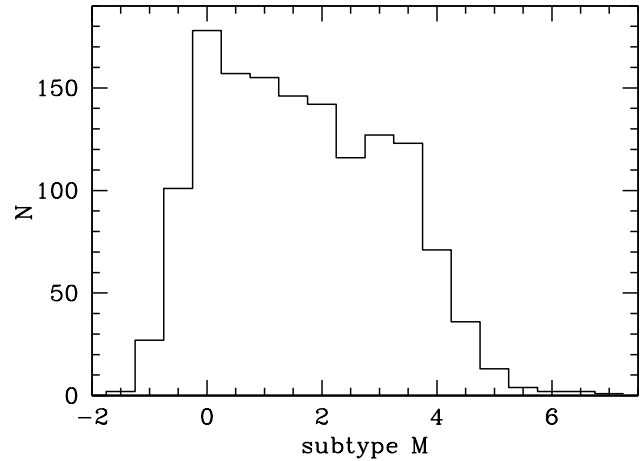}
\caption{Distribution of spectral subtype M, for the stars in our
  survey, with K5=-2 and K7=-1. Most stars are found to be early-M
  objects. The sharp drop at subtypes K7.5 and earlier is a
  consequence of our initial $V-J>2.7$ color cut, which was meant to
  select only stars cool enough and red enough to be M dwarf. The
  sharp drop for subtypes M4 and later is a consequence of the high
  magnitude limit ($J<9$) of our survey, which restricts the distance
  range over which late-M stars are selected. The magnitude limit also
  explains the slow drop in numbers from subtypes M0 to M3, whereas
  one would normally expect the lower-mass M3 stars to be more common
  than earlier-type objects in a volume limited
  sample.\label{sptype_hist}}
\end{figure}

A histogram of the distribution of spectral subtypes is shown in
Figure~\ref{sptype_hist}, with the final tally compiled in
Table~\ref{table_st}. Most of the stars in our survey have
subtypes in the M0.0-M3.0 range. The sharp drop for stars of subtypes
K7.5 and K7.0 is explained by the color selection used in the
\citet{LepineGaidos.2011} catalog ($V-J>2.7$) which was originally
intended to select only M dwarfs; note that stars with subtypes
earlier than K7.0 are also excluded from the graph, and are probably
contaminants of the color selection in any case. Our deficit of K7.0
and K7.5 stars however spectra demonstrates that the adopted selection
criterion is efficient in excluding K dwarfs from the catalog. The
distribution of spectral subtypes also shows a marked drop in numbers
for subtypes M4 and later. This is a consequence of the relatively
bright magnitude limit ($J<9$) of our subsample, combined with the low
absolute magnitudes of late-type M dwarfs, which excludes most
late-type stars from our survey, since these tend to be fainter than
our magnitude limit, even relatively nearby ones. 

\subsection{Semi-automated classification using THE HAMMER}

\begin{figure}
\plotone{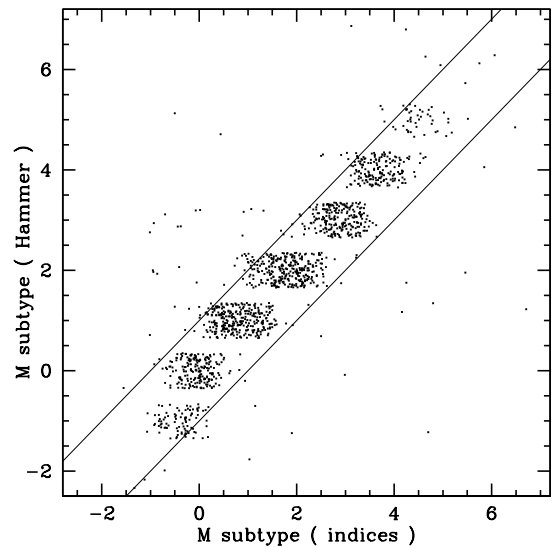}
\caption{Comparison of spectral types assigned by the spectral index
  method and those assigned by the Hammer code, which is used for
  stellar classification in the Sloan Digital Sky Survey. Subtypes
  generally agree to within the advertized precision of the Hammer ($\pm$1
  subtype, illustrated by the slanted lines), although the Hammer
  subtypes are marginally later than the spectral index subtypes, by
  0.27 subtype on average.\label{spec_comp}}
\end{figure}

To verify the accuracy and consistency of spectral typing based the
spectral-index method described above, we performed independent spectral
classification using the Hammer code \citep{Covey2007}. The Hammer was
designed to classify stars in the Sloan Digital Sky Survey
Spectroscopic database, including M dwarfs \citep{West2011}. The code
works by calculating a variety of spectral-type sensitive band
indices, and uses a best-fit algorithm to identify the spectral
subtype providing the best match to those band indices. For late-K and
M dwarfs, spectral subtypes are determined to within integer value
(K5, K7, M0, M1, ..., M9).

However, to ensure that the automatically determined spectral types were
accurate we used the manual ``eye check'' mode of the Hammer (version
1.2.5). This mode is typically used to verify that there are no
incorrectly typed interlopers. The Hammer allows the user to compare
spectra to a suite of template spectra to determine the best
match. \citet{West2011} have found that for late-type M dwarfs, the
automatic classifications were systematically one subtype earlier than
those determined visually. Our analysis confirms this offset, and we
therefore disregarded the automatically determined Hammer values to
adopt the visually determined subtypes. 

The resulting subtypes are listed in Table~\ref{table_spectro}. 
Some 170 stars were not found to be good fits to any of the K5, K7, or
M type templates, and thus identified as early-K or G dwarfs. This
subset includes all of the 156 stars that were visually identified as
non-M dwarfs on first inspection (see \S3.1). The remaining 14 stars
were initially found to be consistent with late-K stars, and
classified as K7.0 and K7.5 objects using the spectral index method
described in \S3.2 above; we investigated further to determine why the
stars were classified as early K using the Hammer. On closer
inspection, we found that 3 of the stars are indeed more consistent
with mid-K dwarfs that K7.0 or K7.5, and we thus overran the spectral
index classification and reclassified them as ''G/K'' in
Table~\ref{table_spectro}. For the other 11 stars flagged as mid-K
type with the Hammer, we determined that the stars do show significant
evidence for TiO absorption, which warrants that the stars retain
their spectral-index classifiction of  K7.0/K7.5. 

In the end Table~\ref{table_spectro} lists 159 stars from our initial
sample that are identified and early-type G and K dwarf contaminants,
and most likely made our target list due to inaccurate or unreliable
$V-J$ colors. The remaining 1405 stars are formally classified as
late-K and M dwarfs.

A comparison of spectral subtypes determined from the spectral-index
and Hammer methods is shown in Figure~\ref{spec_comp}. Because the
Hammer yields only integer subtypes, we have added random values in
the $-0.4,0.4$ range to facilitate the comparion. Slanted lines in
Figure~\ref{spec_comp} show the range expected if the two
classification methods (spectral index, Hammer) agree to within 1.0
subtypes. There is however a mean offset of 0.26 subtypes between the
spectral index and Hammer classifications, with the Hammer subtypes
being on average slighlty later.

Figure~\ref{spec_comp} also reveals a number of outliers with large
differences in spectral subtypes between the two methods. We found 48
stars with differences in spectral subtyping larger than $\pm$1.5. The
spectra from these stars were examined by eye: except for one star, we
found the band-index classifications to agree much better with the
observed spectra than the Hammer-determined subtypes. The one
exception is the star PM I11055+4331 (Gl 412B) which the band index
measurements classify as M6.5; in this case the Hammer determined
subtype of M 5.0 appears to be more accurate. This exception likely
occurs because of the saturation of the CaH2 and CaH3 subtypes in the
late-type star, which make the band-index classification less
reliable.

\subsection{Comparison with the ``Meet the Cool Neighbors'' survey}

\begin{figure}
\plotone{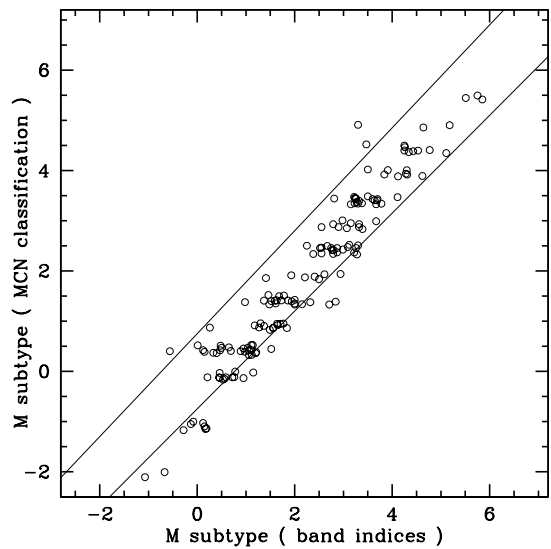}
\caption{Comparison of spectral types assigned by our spectral index
  method and those assigned in the "Meet the Cool Neighbors" (MCN)
  spectroscopic follow-up program. This shows all 161 stars in common
  between our survey and the MCN program. Spectral index subtypes are
  show before rounding up to the nearest half-integer; the MCN
  subtypes have random values of $\pm$0.2 to help in the
  comparison. The classifications generally agree to within $\pm$0.5
  subtypes (slanted lines). Our subtypes are however marginally later
  (by 0.28 subtypes on average) than the MCN
  subtypes.\label{mcn_comp}}
\end{figure}

After the PMSU survey, the largest spectroscopic survey of M dwarfs in
the northern sky was the one presented in the ``Meet the Cool
Neighbors'' (MCN) paper series
\citep{CruzReid.2002,Cruzetal.2003,Reidetal.2003,Reidetal.2004,Cruzetal.2007,Reid2007}. In
this section we compared the spectral clssification from the MCN
survey with our own spectral type assignments. 

To identify the stars in common between the two surveys, we first
performed a cross-correlation of the celestial coordinates of the
stars listed in the MCN tables, to the coordinates listed in the
SUPERBLINK catalog. This was performed for the 1077 stars in MCN which
are north of the celestial equator. We found counterparts in the
SUPERBLINK catalog to within 1\arcsec for 860 of the MCN stars. Of the
217 MCN stars with no obvious SUPERBLINK counterparts, 148 are
classified as ultracool M dwarfs (M7-M9) or L dwarfs (L0-L7.5) which
means they are very likely missing from the SUPERBLINK catalog because
they are fainter than the V=19 completeness limit of the catalog. Of
the 71 remaining stars, close examination of Digitized Sky Survey
scans failed to identify the stars at the locations quoted in MCN. A
closer examination of the fields around those stars identified 49
cases where a high proper motion stars could be found within 3\arcmin
of the quoted MCN positions. These nearby high proper motion stars are
all listed in SUPERBLINK, and have colors consistent with M dwarfs; we
therefore assumed that the quoted MCN positions are in error, and
matched those 49 MCN entries with the close by high proper motions
stars from SUPERBLINK. Of the remaining 22 stars we found 6 that have
proper motions below the SUPERBLINK limit of $\mu>40$ mas yr$^{-1}$
and another 5 stars with proper motions within the
SUPERBLINK limit but that appear to have been missed by the SUPERBLINK
survey. Finally, there were 11 MCN stars that we could not identify at
all on the Digitized Sky Survey images, and we can only assume that
the positions quoted in MCN are too large for proper identification,
and that the stars should be considered ``lost''.

Of the 909 stars in the MCN program with SUPERBLINK counterparts, we
found only 219 which satisfy the magnitude limit ($J<9$) of our
present sample of very bright M dwarfs. Of those, 52 stars have colors
bluer ($V-J<2.7$) than our sample limit; 48 of them are classified
as F and G stars in MCN, consistent with their bluer colors. The other
4 stars are classified as M dwarfs, although they have $V-J<2.7$
according to \citet{LepineGaidos.2011}. We infer that our $V-J$ colors
for those stars are probably underestimated, which suggests that our
color selection may be overlooking a small fraction of nearby M
dwarfs. In addition, we found another 6 stars which have $V$
magnitudes and $V-J$ colors within our survey range, but were rejected
by the additional infrared ($J-H$,$H-K$) color-color cuts used in
\citet{LepineGaidos.2011} to filter out red giants. All 6 stars are
very bright in the infrared, and it appears that at least one of the
$H$ or $K$ magnitudes listed in the 2MASS catalog may be in error,
making the stars appear to have $J-H$ and/or $H-K$ colors more
consistent with giants. Four of the stars are classified as M dwarfs
in MCN, the other two are late-K dwarfs. Overall, this makes a total
of 8 M dwarfs from the MCN census that were overlooked in our
selection out of the MCN subset of $\approx$150 nearby M dwarfs. This
suggests that our color cuts, combined with magnitude measurement
errors, might be missing $\sim5\%$ of the very bright, nearby M dwarfs.

In the end, this leaves only 161 stars in common between the MCN
program and our own spectroscopic survey. The stars are all classified
as late-K and M dwarfs by MCN, with subtypes ranging from K5 to
M5.5. All 159 stars are identified with a flag (''M'') in the last
column of Table~\ref{table_spectro}; we note that 82 of these stars
were also observed as part of the PMSU survey. We compare the spectral
type assignments from both surveys in Figure~\ref{mcn_comp}, where the
M dwarf subtypes from MCN are plotted against the (non-rounded)
subtypes calculated from the spectral-band indices. To ease the
comparison, random values of $\pm$0.2 are added
to the MCN subtypes. Overall, our classifications agree to within
$\pm$0.5 subtypes with the MCN values. The MCN subtypes, however, tend
to be marginally earlier on average, by 0.28 subtypes; this is in
contrast with the Hammer classifications (see above) which tend to be
slightly later than our own. For the MCN subtypes, the effect is more
pronounced for the earlier M dwarfs ($<$M2.5), where the mean offset
is 0.43 subtypes, whereas the mean offset is only 0.09 for the later
stars. 

\begin{figure}
\hspace{1.0cm}
\includegraphics[scale=0.9]{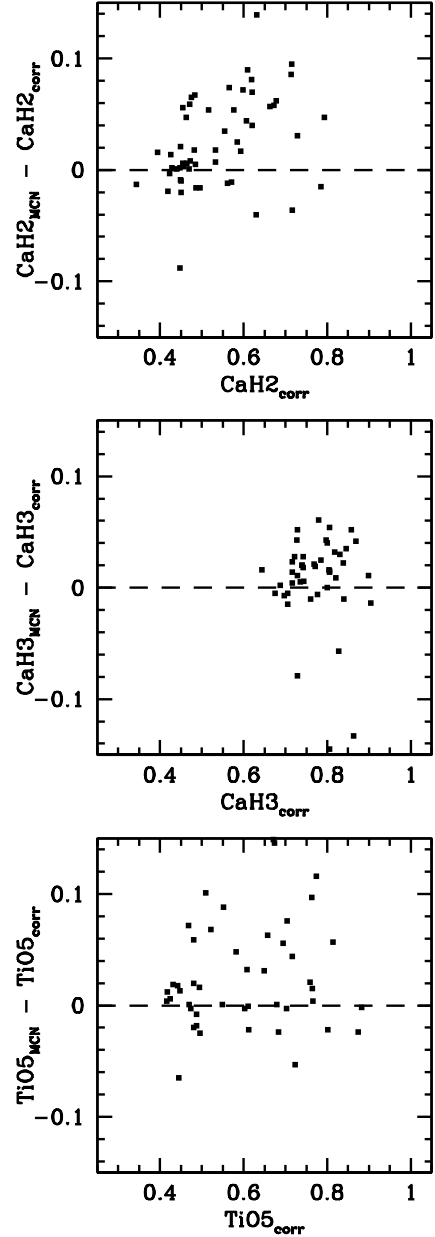}
\caption{Differences between the spectral index values recorded in the
  ``Meet the Cool Neighbors'' (MCN) program and those measured in the
  present survey. The values are compared for a subset of 55 stars in
  common between the two surveys and for which values of CaH2, CaH3,
  and TiO5 were recorded. For the present survey, we use the corrected
  values (CaH2$_c$, CaH3$_c$, and TiO$_c$) as defined in \S3.2.2. The
  MCN values tend to be marginally higher on average, especially for
  stars of earlier spectral subtypes. This explains the marginally
  earlier spectral subtype assigments in MCN compared with the
  ones presented in this paper (see
  Figure~\ref{mcn_comp}).\label{mcn_idx}}
\end{figure}

To investigate the difference in spectral subtype assignments, we
compare the recorded values of the CaH2, CaH3, and TiO5 indices
between the MCN program and our own survey. After a search of the
various tables published in the MCN series of papers, we identified 54
stars in common between the two programs, and for which values of
CaH2, CaH3, and TiO5 were also recorded in both. The differences
between the spectral index values are shown in
Figure~\ref{mcn_idx}. For our own survey, the corrected values of
these indices are used, i.e. CaH2$_c$, CaH3$_c$, and TiO5$_c$ as
defined in \S3.2.2. We find that the CaH2 and TiO5 are estimated
marginally higher in the MCN program than they are in our survey, and
this very likely explains the difference in spectral typing: the 
higher index values yield margnially earlier spectral subtypes. This
again emphasizes the variation in the spectral index measurements due
to spectral resolution and other instrumental setups, and the need to
apply systematic corrections between observatories to obtain a uniform
classification system.

We note that smaller subsets of stars in our census may also have
spectroscopic data published in the literature, from various other
sources. This is especially the case for the 102 stars from the CNS3
and stars with very large proper motions $mu>0.2\arcsec$ yr$^{-1}$
which have been more routinely targeted for follow-up spectroscopic
observations. Additional pectroscopic surveys of selected bright M
dwarfs include, \citet{Scholz2002}, \citet{Scholz2005},
\citet{Reyle2006}, and \citep{Riaz2006}, which all have a few stars in
common with our catalog. Other surveys of nearby M dwarfs have mainly
been targeting fainter stars
\citep{Bochanski2005,Bochanski2010,West2011}, and do not overlap with
our present census.

\subsection{Color/spectral-type relationships}

\begin{figure}
\vspace{-0.2cm}
\hspace{0.1cm}
\includegraphics[scale=0.8]{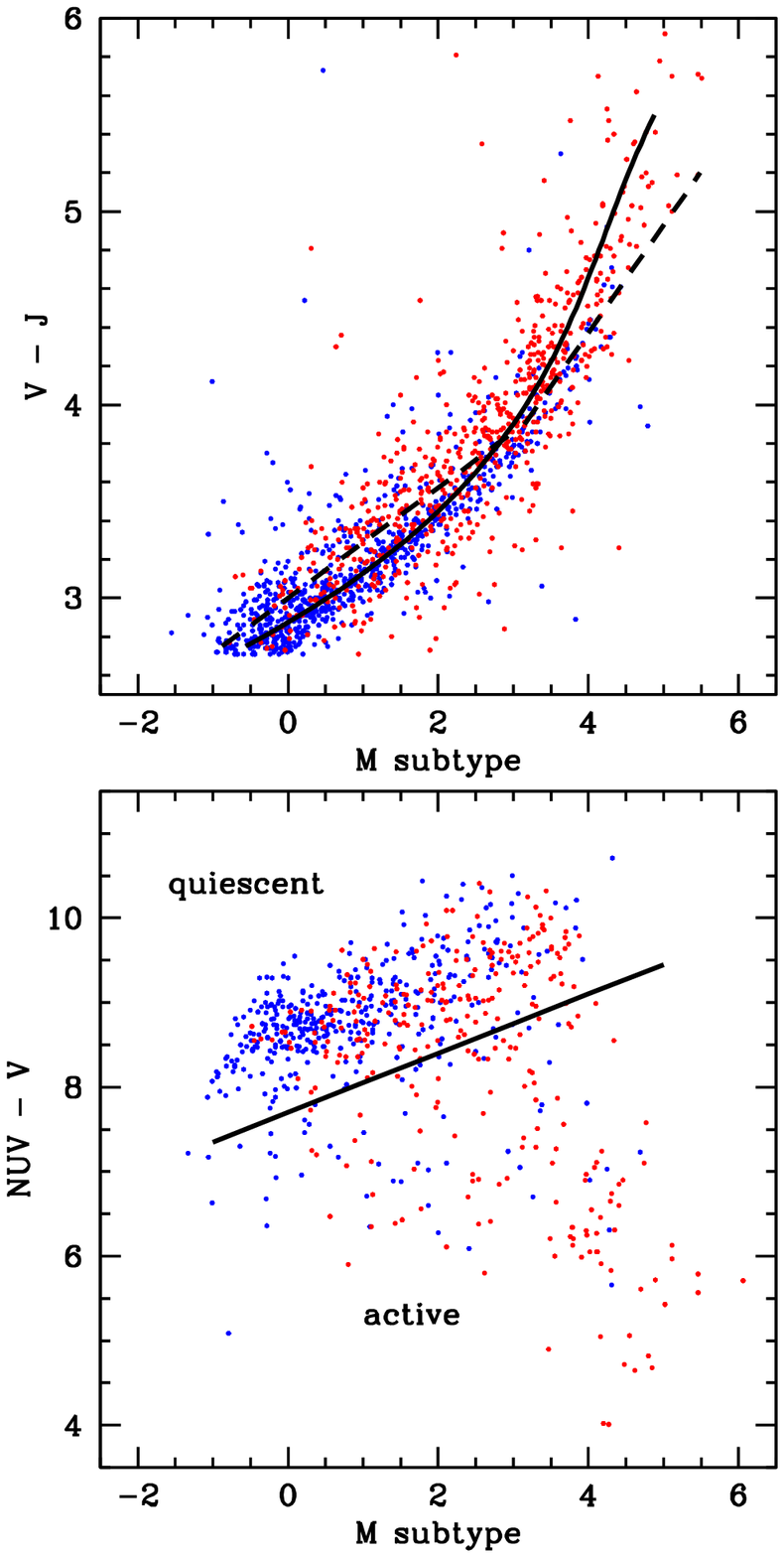}
\caption{Variation of the UV-to-optical $NUV-V$ color index and the
  optical-to-IR $V-J$ color index as a function of spectral subtype
  M. Stars with more reliable V magnitudes from the Tycho-2 catalog
  are shown in blue, stars with V magnitudes derived from less
  reliable photometric measurements are shown in red. The distribution
  of ultra-violet to optical colors ($NUV-V$) with spectral subtype
  shows two populations, one with a tight correlation, consistent with
  blackbody distribution and labeled ``quiescent'', and a scattered
  population of stars with clear $UV$ excess, labeled ``active''. The
  distribution of $V-J$ with subtype closely follows the relationship
  used by \citet{LepineGaidos.2011} to predict subtypes from $V-J$
  colors (dashed line) except for stars of later subtypes which have
  redder colors than predicted. A revised relationship (full line) is
  fitted to the data. Outliers point to stars with bad $V$ magnitude
  measuremens.\label{color_spty}}
\end{figure}

\begin{deluxetable*}{cccccccccc}
\tabletypesize{\scriptsize}
\tablecolumns{8} 
\tablewidth{0pt} 
\tablecaption{Colors and $T_{eff}$ for red dwarfs in our survey as a
  function of spectral subtype.\label{color_subtype}}
\tablehead{
\colhead{subtype} & 
\colhead{n$_{NUV-V}$\tablenotemark{1}} & 
\colhead{$\overline{NUV-V}$\tablenotemark{1}} &
\colhead{$\sigma_{NUV-V}$\tablenotemark{1}} &
\colhead{n$_{V-J}$} & 
\colhead{$\overline{V-J}$} &
\colhead{$\sigma_{V-J}$} &
\colhead{$\overline{T_{eff}}$} &
\colhead{$\sigma_{T_{eff}}$}
}
\startdata 
K 7.0   &  11 &  8.17&  0.21&    25 &  2.90 &  0.31 & 4073 & 98\\
K 7.5   &  47 &  8.60&  0.31&    97 &  2.89 &  0.16 & 3883 & 82\\
M 0.0   &  87 &  8.66&  0.36&   175 &  2.94 &  0.21 & 3762 & 71\\
M 0.5   &  79 &  8.74&  0.30&   159 &  3.11 &  0.34 & 3646 & 48\\
M 1.0   &  76 &  8.89&  0.39&   151 &  3.19 &  0.18 & 3565 & 44\\
M 1.5   &  57 &  9.07&  0.38&   147 &  3.36 &  0.23 & 3564 & 39\\
M 2.0   &  57 &  9.25&  0.47&   139 &  3.52 &  0.34 & 3518 & 57\\
M 2.5   &  48 &  9.45&  0.50&   118 &  3.69 &  0.28 & 3500 & 61\\
M 3.0   &  34 &  9.61&  0.38&   125 &  3.91 &  0.28 & 3423 & 62\\
M 3.5   &  28 &  9.69&  0.33&   124 &  4.17 &  0.33 & 3320 & 66\\
M 4.0   &   5 &  9.72&  0.35&    71 &  4.45 &  0.41 & 3204 & 76\\
M 4.5   &   1 &\nodata&\nodata&  36 &  4.81 &  0.46 & 3119 & 43\\
M 5.0   &   0 &\nodata&\nodata&  12 &  5.23 &  0.50 & 3014 & 61
\enddata
\tablenotetext{1}{Non-active (``quiescent'') red dwarfs only.}
\end{deluxetable*}

Spectral subtypes were inititially estimated in
\citet{LepineGaidos.2011} based on $V-J$ colors alone. Here we verify
this assumption and re-evaluate the color-magnitude relationship for
bright M dwarfs. The $V-J$ color index combines estimated optical ($V$)
magnitudes from the SUPERBLINK catalog to the infrared $J$
magnitudes of their 2MASS counterparts. The SUPEBLINK $V$ magnitudes
are estimated either from the Tycho-2 catalog $V_T$ magnitudes, or
from a combination of the Palomar photographic $B_J$ (IIIaJ), $R_F$
(IIIaF), and $I_N$ (IVn) magnitudes, as described in
\citet{LepineShara.2005}. Values of $V$ are more accurate for the
former ($\approx$0.1mag) than for the latter ($\gtrsim$0.5mag);
Table~\ref{table_photo} indicates the source of the V magnitude.

Mean values and dispersion about the mean of the $V-J$ colors are
listed in Table~\ref{color_subtype}, for each bin of half-integer
subtype; the table also lists how many stars of each type are in each
bin.  The $V-J$ colors of our stars are also plotted as a function of
spectral subtype in Figure~\ref{color_spty} (top panel). The adopted
color-subtype relationship from \citet{LepineGaidos.2011}
is shown as a thick dashed line in Figure~\ref{color_spty}. Stars with
the presumably more reliable Tycho-2 magnitudes are shown in blue,
while stars with photographic $V$ magnitudes are shown in red. Stars
with Tycho-2 $V$ magnitudes appear to have marginally bluer colors
at a given subtype; this however is an effect of the visual
magnitude limit of the Tycho-2 catalog, which includes only the
brightest stars in the $V$ band and is thus more likely to list bluer
objects. The \citet{LepineGaidos.2011} relationship generally follows
the distribution at all subtypes, but with mean offsets up to
$\pm$0.4mag in $V-J$, especially at earlier and later subtypes. We
perform a $\chi^2$ fit to determine the following, improved
relationship:
\begin{displaymath}
\left[ \rm SpTy \right]_{V-J} = {\rm -32.79 + 20.75 ( V - J ) - 4.04 ( V - J )^2}
\end{displaymath}
\begin{equation}
{\rm + 0.275 ( V - J )^3}
\end{equation}
after exclusion of 3-$\sigma$ outliers. The relationship is shown in
Figure~\ref{color_spty} (solid line). There is a scatter of 0.7
subtype between $\left[ SpTy \right]_{V-J}$ and the subtype determined
from spectral band indices. While the spectroscopic classification is
more accurate and reliable, photometrically determined spectral
subtypes using the equation above should still be accurate to $\pm$0.5
subtype about 80\% of the time, and to $\pm$1.0 subtype 95\% of the
time, which may be useful for a quick assessment of subtype when
spectroscopic data is unavailable.

We also compare the near-UV to optical magnitude color $NUV-V$ for the
714 stars in our sample which have counterparts in GALEX; the
distribution is shown in Figure~\ref{color_spty} (bottom panel).
We find that stars become progressively redder as spectral subtype
increases, from $NUV-V=8$ at M0 to $NUV-V=10$ at M4. There is however
a significant fraction of M dwarfs which display much bluer $NUV-V$
colors at any given subtype. The excess in $NUV$ flux is strongly
suggestive of chromospheric activity (see \S6 below for a more
detailed analysis). We separate the active stars from the more
quiescent objects with the following condition:
\begin{equation}
\left[ \rm NUV - V \right] > 7.7 + 0.35 \left[\rm Spty\right] \
\end{equation}
where $\left[\rm Spty \right]$ is the mean spectral subtype calculated from
equations 8-11. After excluding active stars, we calculate the mean
values and scatter about the mean of $NUV-V$ for each half-integer
spectral subtype. Again those are listed in Table~\ref{color_subtype}
for reference; the table also lists the number of non-active stars
used to calculate the mean. There is not a sufficient number of stars
to calculate mean values and scatter at M4.5 (1 star) and M5.0 (0
star).

\section{Survey completeness}

\begin{figure}
\epsscale{1.1}
\plotone{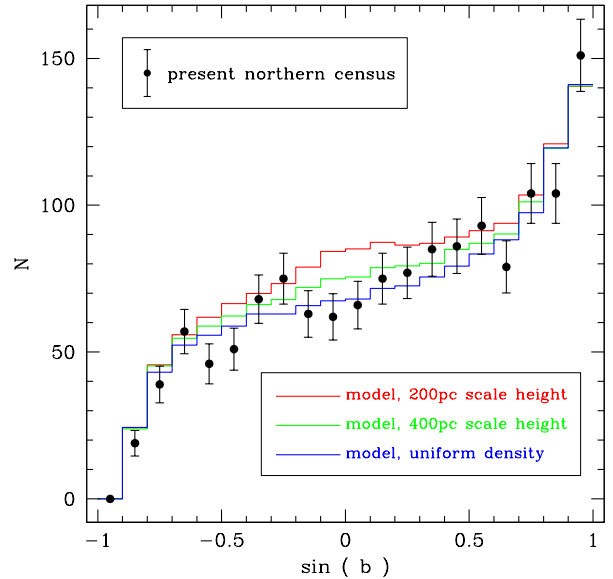}
\caption{Number of spectroscopically confirmed M dwarfs as a function
  of $sin(b)$, where $b$ is the Galactic latitude. Numbers from our
  census are shown as filled circles, with errorbars showing the
  Poisson noise. Distributions predicted from models with either
  uniform space density, or with decreases perpendicular to the
  Galactic plane following scale-heights of 200pc and 400pc, are shown
  for comparison. The observed distribution is largerly
  consistent with a uniform density in the local $d<70$pc
  volume, with no evidence for incompleteness at low Galactic
  latitude.\label{compl}}
\end{figure}

Our 1,564 spectroscopically confirmed M dwarfs are drawn from a
catalog with proper motion limit $\mu<40$ mas yr$^{-1}$. The low
proper motion limit of the SUPERBLINK catalog catches most of the
nearby stars, but potentially overlooks nearby M dwarfs with small
components of motion in the plane of the sky \--- either due to low
space motion relative to the Sun or to projection effects. The catalog
may also be affacted by other sources of incompleteness (e.g. missed
detection, faulty magnitude estimate) which means that at least some
very bright, nearby M dwarfs must be missing from our survey due to
kinematics bias and other effects.

To evaluate the completeness of our census, we first consider the
primary source of incompleteness: the kinematics bias of the proper
motion catalog. As discussed in \citet{LepineGaidos.2011}, the
completeness depends on the local distribution of stellar motions, and
increases with distance from the Sun. To estimate the kinematics bias
in our sample, we built a model reproducing the local distribution and
kinematics of nearby M dwarfs. We first assumed the stars to have a
uniform spatial distribution in the solar vicinity, and generated a
random distribution of $10^{5}$ objects within a sphere of radius
$d=70$pc centered on the Sun. We assigned transverse motions to all the
stars, assuming a velocity-space distribution similar to that of the
nearby (d$<$100pc) G dwarfs in the Hipparcos catalog
\citep{vanLeeuwen2007}. Because the distribution of stellar velocities
is not isotropic, we assigned transverse motions for each simulated star
based on the statistical distribution of transverse motions for
Hipparcos stars with sky coordinates within $30^{deg}$ of the
simulated object. We also used the simplifying assumption that the
local M dwarf population has a uniform distribution of absolute
magnitudes over the range $5<M_J<15$, which is the approximate range
of absolute magnitudes reported in the literature for M dwarfs.

We then counted the total number of stars in the simulation with
apparent magnitudes $J<9$, and calculated the fraction of those stars
with proper motions $\mu>40$ mas yr$^{-1}$. We found that 93\% of
nearby M dwarfs with $J<9$, on average, have proper motions above the
SUPERBLINK limit. M dwarfs with $J<9$ extend to a maximum distance of
63 parsecs; most of the stars which fail the proper motion cut are in
the higher distance range (d$<$50pc), and are stars near the bright
end of the luminosity distribution $M_J\approx5-6$. For this reason,
if we assume a luminosity function which increases at fainter absolute
magnitudes, the fraction of $J<9$ stars which fall within the proper
motion cut is increased, because more of the $J<9$ stars in the local
population are now M dwarfs of lower luminosities and closer
distances, which are more likely to have high proper motions. The
observed field M dwarf luminosity function does indeed increase for
early-type dwarfs, to reach a peak at $M_J\simeq8.0$
\citep{Reid2002,Bochanski2010}, which means that the 93\% completeness
estimated above must be a lower limit. To verify this, we tested a
luminosity function where the number of stars increases linearly with
absolute magnitude and doubles from $M_J=5$ to $M_J=8$; with this
model, our simulations showed that 96\% of all $J<9$ M dwarfs, would
have $\mu>40$ mas yr$^{-1}$, and thus be within the detection limit of
SUPERBLINK. Overall, this suggests that only about $5\%$ of all $J<9$
M dwarfs on the sky will be overlooked in our census because of the
proper motion bias.

The SUPERBLINK surveys does however suffer from various other sources
of incompleteness, such as the inability to detect moving stars in
saturated regions of photographic plates (i.e. in the immediate
vicinity of very bright stars), or a difficulty in detecting stars in
very crowded field. Also, the SUPERBLINK code has trouble detecting
the motions of relatively bright $V<12$ stars because of the saturated
cores of their point spread functions. In practice, this is mitigated
by incorporating data from the Tycho-2 catalog, which provides very
accurate proper motion measurements for bright stars. However the
Tycho-2 catalog itself has some level of incompleteness in the
$8<V<12$ manitude range. 

One way to test incompleteness due to crowding or saturation is to
examine the distribution of SUPERBLINK stars as a function of Galactic
latitude. Crowding and saturation effects should be more pronounced in
low Galactic-latitude fields, where the stellar density is high. In
comparison, stars in high Galactic latitude fields will be easier to
detect with the SUPERBLINK code. The same is true for the Tycho-2
catalog, which should be more complete at high Galactic latitudes.

We calculated the number of spectroscopically confirmed M dwarfs in our
sample as a function of $sin(b)$, where $b$ is the Galactic
latitude. The distribution is shown in Figure~\ref{compl}; the number
of stars in each bin is plotted as a filled circle, with errobars
showing the $1\sigma$ Poisson error. The distribution increases with
$sin(b)$ because our stars are all located north of the celestial
equator. For comparison, we plot the trend expected of a uniform
distribution of stars in the local volume (blue histogram), assuming
the same total number of stars as in our census. We find our data to
be largely consistent with the uniform distribution model, and see no
evidence of a significant dip at low Galactic latitude, which one
would expect if the SUPERBLINK survey is incomplete at low $b$. This
suggests that our sample does not suffer from significant sources of
incompleteness due to saturation/crowding.

One might however argue that the uniform density assumption is invalid
for a census extending to $\simeq$70pc of the Sun, and that a slight
overdensity of stars should be detected at low Galactic latitudes. The
fact that we see no evidence of such an overdensity in our data would
then be indirect evidence of incompleteness at low Galactic
latitude. To examine this possibility, we estimated the expected
distributions from simulations in which the stellar density
perpendicular to the plane (i.e. along $Z$) decreases with scale
heights of 400pc or 200pc. Kinematics and absolute magnitude
distributions were
also simulated as described above, and stars were selected based on
$J<9$ and $\mu>40$ mas yr$^{-1}$. The mean distributions from the
400pc and 200pc scale-height models are shown in Figure~\ref{compl},
and do indeed predict a slight overdensity of objects at low Galactic
latitudes. Assuming that all the stars at high Galactic latitude are
detected in all the models, we find a $\sim$5\% excess of stars in the
400pc scale-height model over the observed number, and a $\sim$10\%
excess of stars in the 200pc scale-height model. One might then argue
that the SUPEBLINK census potentially has an additional 5\%-10\%
incompleteness level, due to incompleteness in the plane of the Milky
way. This is most probably an overestimate, however, because there is
no evidence that the local stellar distribution shows any significant
decrease with $Z$. Overall, the SUPERBLINK census appears to be
essentially complete at low Galactic latitudes.

Proper-motion selection may introduce an additional bias against
metal-rich and/or young stars, which tend to have lower components of
motion in the vicinity of the Sun. However this effect is expected to
be small, e.g. a few percent against [Fe/H] = 0 relative to [Fe/H] =
-0.5 at 45~pc \citep{Gaidos2012}. Nearly all likely members of the
Hyades \citep{Perryman1998}, Ursa Majoris \citep{King2003}, and TW
Hydrae \citep{Reid2003} young nearby moving groups have proper motions
that exceed 40~mas yr$^{-1}$ and thus would not be selected against using
the current selection method.

Finally, some bright M dwarfs may be overlooked because of the imposed
color cuts, due to magnitude errors and uncertainties. As described in
\S3.4 above, a handful of previously known M dwarfs from the MCN
census were overlooked in our target selection for precisely those
reasons. As suggested in \S3.4, it is possible that we may be
overlooking 5\% of bright M dwarfs because of magnitude errors. This,
combined with the estimated 96\% completeness in the proper motion
selection estimated above, suggests that our list of 1405 M dwarfs
likely includes $\approx91\%$ of all existing M dwarfs with infrared
magnitude $J<9$ as seen from Earth. There may still be $\approx$140
bright M dwarfs to be identified, although the precise number can only
be determined after these ``missing'' stars are found.  

\section{Phoenix Model fits and $T_{eff}$ estimates}

\begin{figure}
\epsscale{1.2}
\plotone{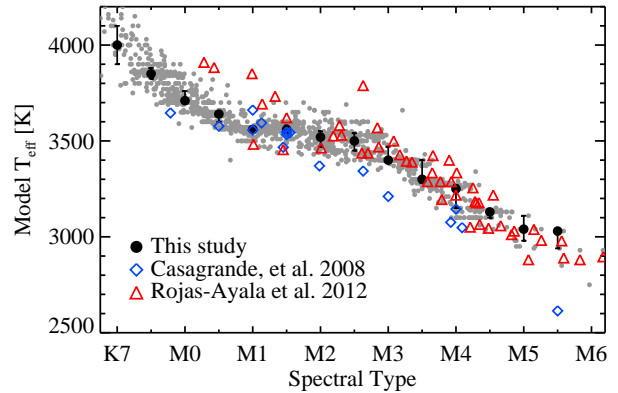}
\caption{Effective temperatures for the bright M dwarfs in our survey
  determined by fits to the PHOENIX models. The gray points represent
  individual objects, while the large black points are the median
  values of the objects within each half subtype bin. The error bars
  are the interquartile ranges. Blue diamonds and red triangles
  show the T$_{eff}$ estimates for subsets of stars in our survey
  whose tempreatures were estimated from photometry
  \citet{Casagrande2008} and model fits to infrared spectra
  \citep{Rojas-Ayala2012} respectively. While both spectroscopic
  estimates show evidence of a plateau at M2-M3, the photometric
  estimates do not concur. \label{spt_color}}
\end{figure}

We compared our spectra to a grid of 298 models of K- and M-dwarf
spectra generated by the BT-SETTL version of PHOENIX \citep{Allard2010}.
BT-SETTL includes updated opacities (i.e. of H$_2$O), revised solar
abundances \citep{Asplund2009}, a refractory cloud model, and
rotational hydrodynamic mixing.  The models include effective
temperatures $T_{eff}$ of 3000-5000~K in steps of 100~K, $\log g$ values
of 4, 4.5, and 5, and metallicities of [M/H] = -1.5, -1, -0.5, 0,
+0.3, and +0.5. For each temperature, $\log g$ and metallicity value,
we selected the model with $\alpha$/Fe that was closest to solar. 

\begin{figure*}
\plotone{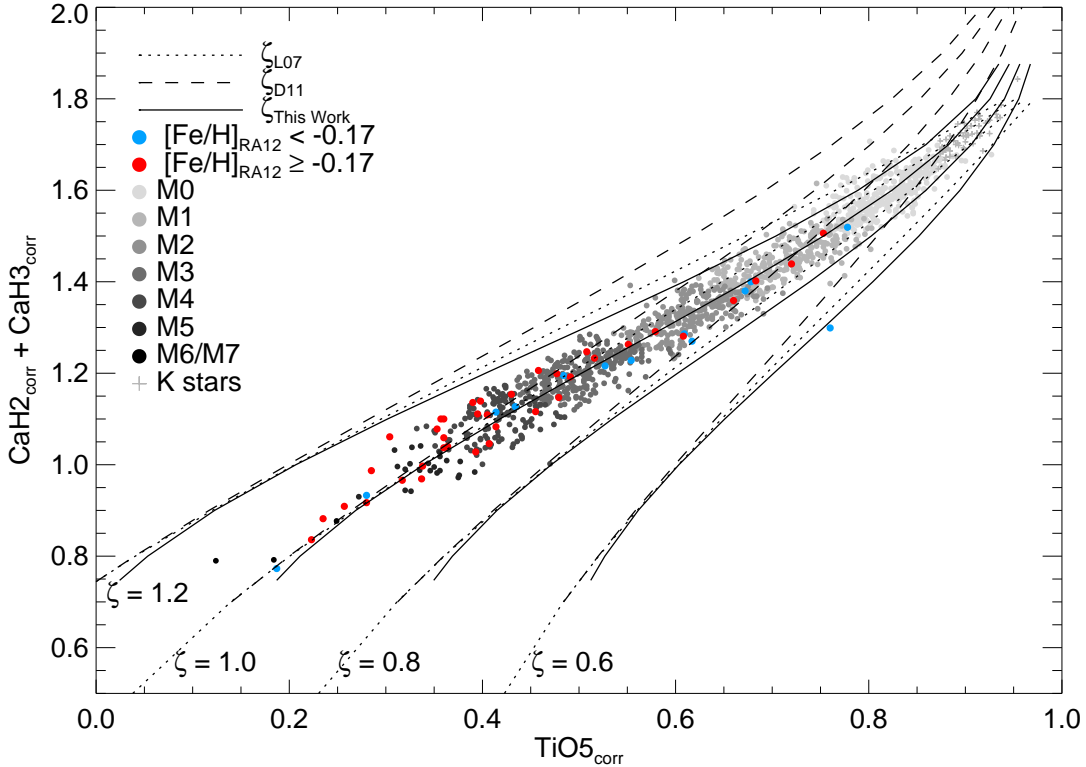}
\caption{Distribution of corrected CaH2+CaH3 vs TiO5 band index
  values for the stars in our survey (grey dots - with brightness
  levels correlating with spectral subtype). There is a tight correlation
  between the two indices, which are both also correlated
  with spectral subtypes, with earlier stars on the upper right of the
  diagram as shown. The distribution is used as a guide to calibrate
  the value of $\zeta$, with $\zeta=1$ assumed to trace the
  CaH2+CaH3/TiO5 relationship for stars with average Galactic disk
  abundances (near-solar). The iso-$\zeta$ contours from the earlier
  calibrations of \cite{Lepine2007} and \cite{Dhital2012} are shown as
  dotted and dashed lines, respectively. When applied to our corrected
  spectral index values, they both diverge from the observed
  distribution at earlier subtypes, with the \cite{Lepine2007}
  overestimating the $\zeta$ of late-K and early-M dwarfs, while the
  \cite{Dhital2012} calibration yield underestimates. This emphasizes
  again the need to use properly recalibrated and corrected spectral
  index values (see Figures~\ref{pmscomp}-\ref{idx_comp}). Our
  revised, dataset-specific calibration is shown with the continuous
  lines. Known metal-rich amd metal-poor stars are denoted in red and
  blue, respectively.\label{cah_tio}}
\end{figure*}

The spectral density of model calculations varies with wavelength but
is everywhere vastly greater than the resolution of our spectra.
Model spectra were thus convolved with a gaussian with FHWM of the same
resolution as the spectra, and a corrective shift (typically less than
a resolution element) was found by cross-correlating the observed and
model spectra.  Normalized spectra were ratioed and $\chi^2$
calculated using the variance spectrum of the observations.  We
restricted the spectral range over which $\chi^2$ is calculated to
5600-9000\AA\ and excluded the problematic region 6400-6600\AA\ which
contains poorly-modeled TiO absorption \citep{Reyle2011}. We also
excluded regions where the telluric correction is rapidly changing with
wavelength, i.e. the slope, smoothed over 4 resolution elements or
11.7\AA, exceeds 1.37 \AA$^{-1}$. The model with the smallest $\chi^2$
value was identified. For a more refined estimate of effective temperature,
we selected the 7 best-fit models and constructed 10,000 linear
combinations of them; the ``effective temperature'' of each is the
weighted sum of the temperatures of the components.  We again found
the model with the minimum $\chi^2$.  We calculated the standard
deviation of $T_{eff}$ among the combination models as a function of the
maximum allowed $\chi^2$.  We reported the maximum standard deviation
as a conservative estimate of uncertainty.  We also calculated formal
95\% confidence intervals for $T_{eff}$ based solely on the expected
distribution of $\chi^2$ for $N-3$ degrees of freedom, where $N \sim
1100$ is the number of resolution elements used in the fit.  The
parameters of the best-fit model, and the refined $T_{eff}$, standard
deviation, and confidence intervals are reported in
Table~\ref{table_spectro}.
 
Values of $T_{eff}$ calculated for individual stars are plotted in
Figure~\ref{spt_color} as a function of their spectral subtype (grey
dots). Median values for stars within each half-subtype bin are
plotted in black, with error bars showing the interquartile
ranges. Our model-fit algorithm prefers values that match the $T_{eff}$
model grid, which have a 100K grid step (i.e., 3500K is preferred over
3510K).

Our results suggest the existence of a $T_{eff}$ plateau spanning
M1-M3. To investigate this further, we compare our values to the
effective temperatures reported in \citet{Casagrande2008} for 18 of
the stars in our sample, and in \citep{Rojas-Ayala2012} for another 49
stars; our own spectral type determinations are combined to the
T$_{eff}$ measured by the other authors. The values are compared in
Figure~\ref{spt_color}. We find that our mid-type plateau is
corroborated with the \citep{Rojas-Ayala2012} values, but not with
those from \citet{Casagrande2008}, whose values decrease more linearly
with spectral subtype. The effective temperatures in
\citet{Casagrande2008} are based on photometric measurements while
the \citep{Rojas-Ayala2012} values are estimated by PHOENIX model fits
to infrared spectra. It is interesting that the fits to the optical
and infrared spectra yield T$_{eff}$ values which are in
agreement. The disagreement with the photometric determinations
however suggest that atmospheric models for M dwarfs are still not
well understood, in particular in the M1-M4 spectral regimes.

\section{The $\zeta$-Parameter and Metallicity Estimates}

\begin{figure}
\epsscale{2.2}
\plotone{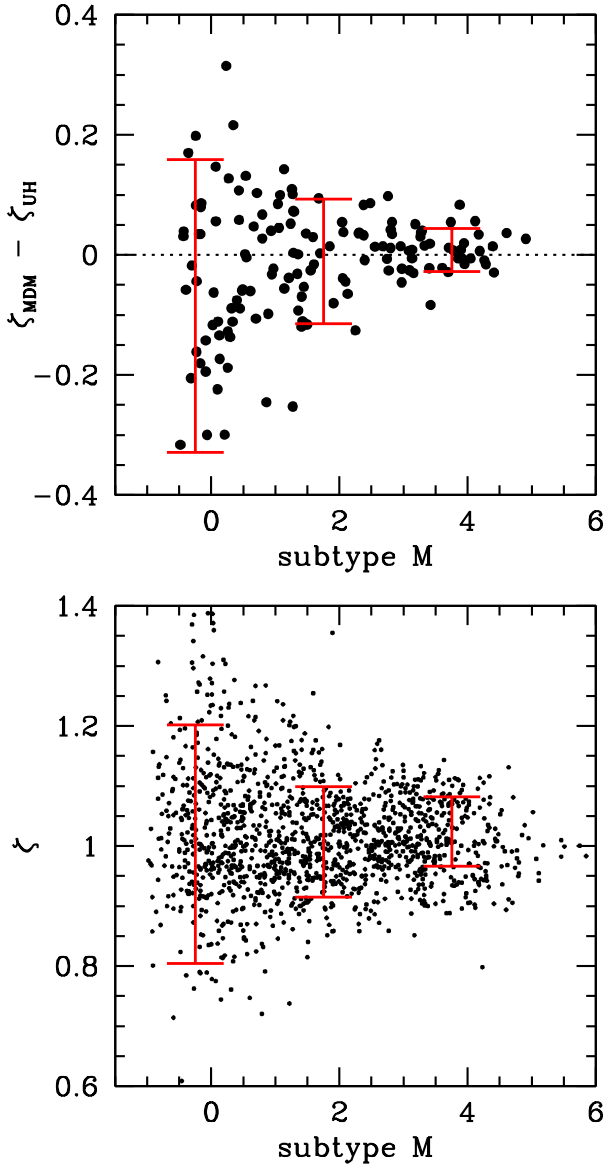}
\caption{The $\zeta$ parameter as a function of spectral subtype. Top:
  difference in $\zeta$ for the same stars measured at the two
  observatories (MDM and UH). The scatter provides an estimate of the
  measurement error on $\zeta$, which is significantly larger at
  earlier subtypes. Bottom: adopted values of $\zeta$ for all the
  stars in the survey. The larger scatter at subtypes M2 and earlier
  can be fully accounted by the measurement errors. The scatter at
  subtypes M3 and later is larger than the measurement error, and is
  thus probably intrinsic and is evidence for intrinsic metallicity
  scatter in the solar neighborhood.\label{zeta_comp}}
\end{figure}

\subsection{Recalibration of the $\zeta$ parameter}

The $\zeta_{TiO/CaH}$ parameter (denoted $\zeta$ for short) is a
combination of the TiO5, CaH2, and CaH3 spectral indices which was
shown to be correlated with metallicity in metal-poor M subdwarfs
\citep{Woolf2009}. The index was first described in
\citet{Lepine2007}, and a revised calibration has recently been
proposed by \citet{Dhital2012}. The index measures the relative
strength of the TiO molecular band around 7,000\AA\ with respect to the
nearby CaH molecular band. In cool stars, in fact, the ratio between
TiO and CaH is a function of both gravity and metallicity. The CaH
band is noticeably stronger in giants \citep{Mann2012}, and this
effect can be used as affective means to separate out M giants from M
dwarfs using optical spectroscopy. In the higher gravity M
dwarfs/subwarfs however, the TiO to CaH ratio is however believed to
be mostly affected by metallicity, although young stars may show
gravity effects as well. 

The $\zeta$ parameter was originally introduced to rank metal-poor,
main-sequence M stars into three metallicity classes
\citep{Lepine2007}; stars with $0.5<\zeta<0.825$ are formally
classified as subdwarfs (sdM), $0.2<\zeta<0.5$ defines extreme
subdwarfs (esdM), while a $\zeta<0.2$ identifies the star as an
ultrasubdwarf (usdM). However, it is conjectured that $\zeta$ could
be used to measure metallicity differences in disk M dwarfs, i.e. at
the metal-rich end. Disk M dwarfs are generally found to have
$0.9<\zeta<1.1$, though it is unclear if variations in $\zeta$
correlate with metallicity for values within that range. Measurement
of Fe lines in a subset of M dwarfs and subdwarfs does confirm that
the $\zeta$ parameter is correlated with metallicity
\citep{Woolf2009}, with $\zeta\simeq1.05$ presumably corresponding to
solar abundances. However there is a significant scatter in the
relationship which raises doubts about the accuracy of $\zeta$ as a
metallicity diagnostic tool. 

A important caveat is that the TiO/CaH ratio is not sensitive to the
classical iron-to-hydrogen ratio Fe/H, but rather depends on the
relative abundance of $\alpha$-elements to hydrogen ($\alpha$/H)
because O, Ca, and Ti are all $\alpha$-elements. Variations in
$\alpha$/Fe would thus weaken the correlation between $\zeta$ and
Fe/H. The $\alpha$/Fe abundance ratio is however relatively small in
metal-rich stars of the thin disk ($\pm$0.05dex) disk stars, and are
found to be significant ($\pm$0.2dex) mostly in more metal-poor stars
associated with the thick disk and halo \citep{Navarro_etal.2011}. It
is thus unclear whether typical $\alpha$/Fe variations would affect
the $\zeta$ parameter significantly in our subset, which is dominated
by relatively metal-rich stars.

On the other hand, it is clear that the index has significantly more
leverage at later subtypes. This is because the strengths of both the
TiO and CaH bands are generally greater, and their ratio can thus be
measured with higher accuracy. The index is much leass reliable at
earlier M subtypes however, and is notably inefficient for late-K
stars.

A more important issue is the specific calibration adopted for the
$\zeta$ parameter, which is a complicated function of the TiO5, CaH2,
and CaH3 indices. The $\zeta$ parameter itself is defined as:
\begin{equation}
\zeta = \frac{1-{\rm TiO5}}{1-[{\rm TiO5}]_{Z_{\odot}}},
\end{equation}
which in turns depend on $[{\rm TiO5}]_{Z_{\odot}}$, itself a function
of CaH2+CaH3. The function $[{\rm TiO5}]_{Z_{\odot}}$ represents the
expected value of the TiO5 index in stars of solar metallicity, for a
given value of CaH2+CaH3. In \citep{Lepine2007}, ${\rm
  TiO5}]_{Z_{\odot}}$ was defined as:
\begin{displaymath}
[{\rm TiO5}]_{Z_{\odot}} = -0.050 - 0.118 \ {\rm CaH} + 0.670 \ {\rm CaH}^2 
\end{displaymath}
\begin{equation}
- 0.164 \ {\rm CaH}^3
\end{equation}
where ${\rm CaH}={\rm CaH2}+{\rm CaH3}$. The more recent calibration
of \citet{Dhital2012}, on the other hand, uses:
\begin{displaymath}
[{\rm TiO5}]_{Z_{\odot}} = -0.047 - 0.127 \ {\rm CaH} + 0.694 \ {\rm CaH}^2 
\end{displaymath}
\begin{equation}
- 0.183 \ {\rm CaH}^3 - 0.005 \ {\rm CaH}^4 .
\end{equation}
The difference between the two calibrations is mainly in the treatment
of late-K and early-type M dwarfs, as illustrated in
Figure~\ref{cah_tio}. When overlaid on the distribution of CaH2+CaH3
and TiO5 values from our current survey, however, the two calibrations
fail to properly fit the distribution of data points at the earliest
subtypes (high values of CaH2+CaH3 and TiO5). This results in the
\citet{Lepine2007} overestimating the metallicity at earlier subtypes,
while the \citet{Dhital2012} calibration tends to underestimate
metallicity.

In any case, the evidence presented in \S3.3 and \S3.4 which shows
that differences in spectral resolution and flux calibration can yield
differences in the TiO5, CaH2, and CaH3 spectral indices of the same
stars, also suggests that a calibration of the $\zeta$ parameter may
only be valid for data from a particular observatory/instrument. A
general calibration of $\zeta$ may only be adopted after corrections
have been applied as described in Section 3.2. Because we do not have
any star in common with the \citet{Dhital2012} subsample, we cannot
verify the consistency of their $\zeta$ calibration to our data at
this time. In addition, because we have now applied a correction to
our MDM spectral index measurements, the \citet{Lepine2007}
calibration of $\zeta$ may now be off, and should not be used for our
sample.

Instead, we recalibrate the $\zeta$ parameter again, using our
corrected spectral index values. Our fit of $\left[\rm
TiO5\right]_{c}$ as a function of $\left[{\rm
 CaH}\right]_{c}=\left[{\rm     CaH2}\right]_{c}+\left[{\rm
 CaH3}\right]_{c}$ yields:
\begin{displaymath}
[{\rm TiO5}]_{Z_{\odot}} = 0.622 - 1.906 \  (\left[{\rm CaH}\right]_{c}) -
2.211 \ (\left[{\rm CaH}\right]_{c})^2 
\end{displaymath}
\begin{equation}
- 0.588 \ (\left[{\rm CaH}\right]_{c})^3.
\end{equation}
We calculate the new $\zeta$ values using the corrected values of the
TiO5 index, i.e.:
\begin{equation}
\zeta = \frac{1-{\rm \left[TiO5\right]_{c}}}{1-[{\rm TiO5}]_{Z_{\odot}}}.
\end{equation}
All our values of $\zeta$ are listed in Table~\ref{table_spectro}.

In order to evaluate the accuracy of the $\zeta$ measurements, we
compared the values of $\zeta$ independently measured at both MDM and
UH for the 146 stars in our inter-observatory subset. Values are
compared in Figure \ref{zeta_comp} (top panel) which shows $\Delta
\zeta=\zeta_{MDM}-\zeta_{UH}$ as a function of spectral subtype. We
find a mean offset $\bar{\Delta\zeta}=-0.01$ and a
dispersion $\sigma_{\Delta\zeta}=0.10$. The small offset indicates
that the $\zeta$ measurements are generally consistent between the two
observatories. The dispersion $\sigma_{\Delta\zeta}$, on the other
hand, provides an estimate of the measurement accuracy. Splitting the
stars in three groups, we find the mean offsets and dispersions
$(\bar{\Delta\zeta},\sigma_{\Delta\zeta})$ to be (-0.086,0.244) for
subtypes K7.0-M0.5, (-0.011,0.103) for subtypes M1.0-M2.5, and
(0.008,0.036) for subtypes M3.0-M5.5. Assuming that stars do not show
significant variability in those bands, we adopt the dispersions as
estimates of the measurement errors on $\zeta$ for that particular
subtype range. It is clear from Figure~\ref{cah_tio} that early type
stars should have larger uncertainties in $\zeta$ because of the
convergence of the iso-$\zeta$ lines. The best leverage for estimating
metallicities from the TiO and CaH bandheads is at later types when
the molecular bands and well developed.

The overall distribution of $\zeta$ values as a function of spectral
subtype also shows a decrease in the dispersion as a function of
spectral type (Figure~\ref{zeta_comp}, bottom panel). In early-type
dwarfs (K7.0-M0.5), the scatter in the $\zeta$ values is relatively
large, with $\sigma_{\zeta}\simeq0.174$. It then drops to
$\sigma_{\zeta}\simeq0.100$ for subtypes M1.0-M2.5, and to
$\sigma_{\zeta}\simeq0.059$ for subtypes M3.0-M5.5. Note that the
scatter in the M3.0-M5.5 bin is a factor 2 larger than the estimated
accuracy of the $\zeta$ for that range, as estimated above. We
suggests this to be evidence of an intrinsic scatter in the $\zeta$
values for the stars in our sample, which we allege to be the
signature of a metallicity scatter. If we subtract in quadrature the
$0.035$ measurement error on $\zeta$, we estimate the instrinsic
scatter to be $\approx$0.05 units in $\zeta$. This intrinsic scatter,
which presumably affects all subtypes equally, is unfortunately
drowned in the measurement error at earlier subtypes ($<$M3).


\subsection{Comparison with other metallicity diagnostics}

\begin{figure}
\vspace{-0.6cm}
\hspace{-0.5cm}
\includegraphics[scale=0.9]{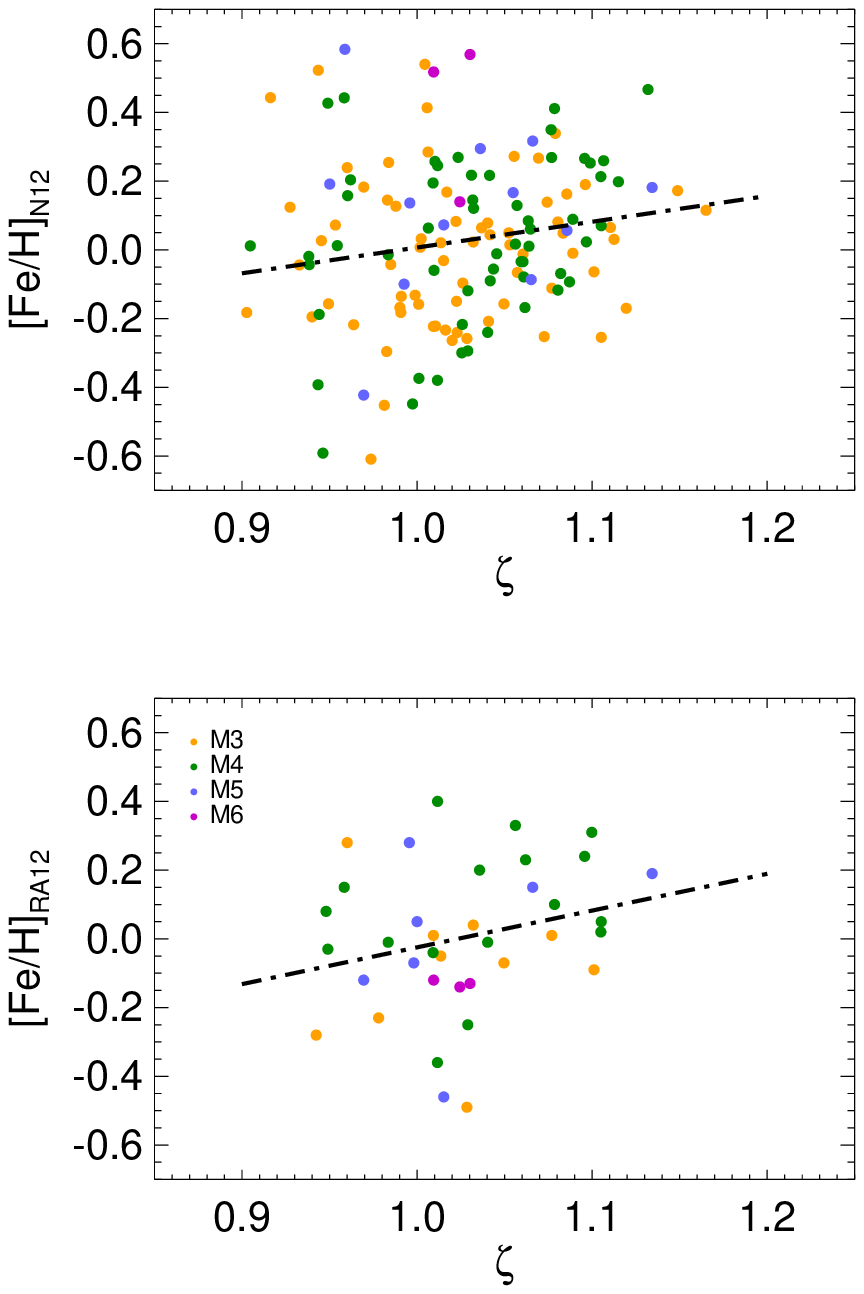}
\caption{Comparison between the $\zeta$ parameter values and
  independent metallicity measurements for subsets of M dwarfs in our
  survey. Top panel compares our $\zeta$ to the Fe/H estimated from
  the (V-K,M$_K$) calibration of \citet{Neves2012} for the same
  stars. Bottom panel compares our $\zeta$ values to the Fe/H
  estimated from the infrared K-band index by
  \citet{Rojas-Ayala2012}. Both distribution show weak positive
  correlations.\label{zeta_feh}}
\end{figure}

\begin{figure*}
\plotone{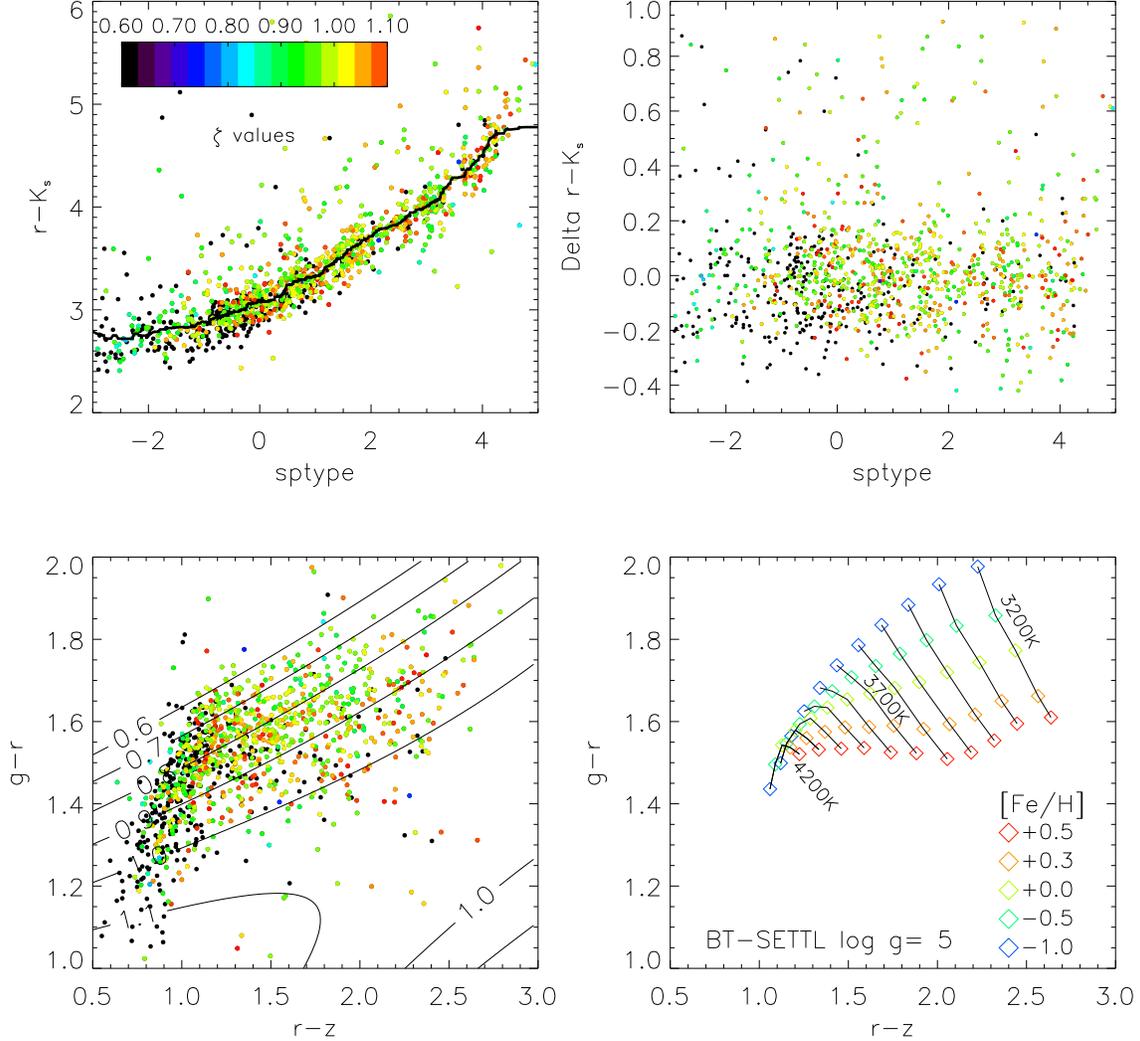}
\caption{SDSS photometry of M dwarf stars synthesized from SNIFS
  spectra and transmission functions convolved with unit airmass.
  Upper left: r-Ks (where Ks is from the 2MASS point source catalog)
  against M spectral subtype (where K7=-1 and K5=-2). Points are
  colored by the $\zeta$ parameter, which measures TiO/CaH ratio and
  is a metallicity diagnostic in the optical. The $\zeta$ values are
  undefined for late K stars, which are plotted as black points. The
  black curve is a running median (N=81). Upper right: difference of
  r-Ks with respect to the running median vs. spectral type, showing
  no obvious correlation with zeta. Lower left: g-r vs. r-z, showing
  an apparent correlation between these colors and $\zeta$. The
  contours are the empirical function for $\zeta$ derived by
  \citet{West2011}. Lower right: SDSS g-r vs. r-z colors generated by
  the PHOENIX/BT-SETTL model \citep{Allard} for log g=5,
  Teff=3500K-4200K, and five different values of the metallicity as
  noted in the legend. The model predicts the more metal-rich M dwarfs
  to have {\em bluer} g-r colors, while being redder in r-z. The color
  dependence on metallicity in most pronounced in late-type stars, and
  nearly vanishes at K7/M0.\label{model}}
\end{figure*}

To test our $\zeta$ as a tracer of metallicity for dM stars from M3 to
M6, we compared the values to two recent [Fe/H] calibration techniques
for M dwarfs with solar metallicities. First, we used the photometric
calibration of \citet{Neves2012}, which is based on
optical-to-infrared V-K color and absolute magnitude M$_{K}$. The
method is sensitive to small variations in V-K/M$_{K}$ and thus
requires an accurate, geometric parallax. A total of 143 stars in our sample have
parallaxes, and thus can have their metallicities estimated with the
method. Figure~\ref{zeta_feh} (top panel) plots the estimated $\rm
[Fe/H]$ as a function of $\zeta$ for those 143 stars. The distribution
shows significant scatter, but we find a weak correlation of $\rm
[Fe/H]$ with $\zeta$, which we fit with the relationship:
\begin{equation}
[Fe/H]_{\rm N12} = 0.750 \zeta - 0.743.
\end{equation}
Stars are scattered about this relationship with a 1-$\sigma$
dispersion of 0.383 dex. One drawback of the photometric metallicity
determination is that it assumes the star to be single. Unresolved
double stars appear overluminous at a given color, and will thus be
determined to be metal-rich. Also, young and active stars often appear
overluminous in the color-magnitude diagram \citep{Hawley2002}, and
their metalicities based on V-K/M$_{K}$ would also be
overestimated. Multiplicity and activity could therefore contribute in
the observed scatter. Stars with $[Fe/H]_{\rm N12}>0.4$, in
particular, could be overluminous in the V-K/M$_{K}$ diagram, as their
$\zeta$ does not suggest them to be metal-rich.

Next, we retrieve metallicity measurements from
\citet{Rojas-Ayala2012}, who estimated $[Fe/H]$ based on the
spectroscopic calibration from infrared K-band atomic features. Their
list has 37 stars in common with our survey. The $[Fe/H]$ values are
plotted as a function of our $\zeta$ values in the bottom panel of
Figure~\ref{zeta_feh}. Again there is significant scatter, but we also
find a weak correlation which we fit with the relationship:
\begin{equation}
[Fe/H]_{\rm RA12} = 1.071 \zeta - 1.096,
\end{equation}
about which there is a dispersion of 0.654 dex. The statistics are
relativey poor at this time, and more metallicity measurements in
the infrared bands would be useful.


The weak correlation found in both distribution is interesting in
itself. Using a sample of stars spanning a wide range of metallicities
and $\zeta$ values, including metal-poor M subdwarfs, extreme
subdwarfs (esdM), and extremely metal-poor ultrasubdwarfs
(usdM), \citet{Woolf2009} determined a relationship of the form
$[Fe/H] = -1.685 + 1.632 \zeta$, over the range $0.05<\zeta<1.10$. All
the stars in the two distributions from the present survey have
$\zeta$ values between $\sim$0.9 and $\sim$1.2, and thus represent the
metal-rich end of the distribution. The weaker slopes we find in our
correlations (0.75 and 1.07) may indicate that the relationship levels
off at high metallicity end, which would make $\zeta$ much less useful
as a metallicity diagnostic tool in Solar-metallicity and metal-rich M
dwarfs. The correlations are however weak, and more accurate
measurement of $\zeta$ and $Fe/H$ would be needed to verify this
conjecture.

\subsection{The one M subdwarf: PM I20050+5426 (V1513 Cyg)}

The primary purpose of the $\zeta$ parameter is the identification of
metal-poor M subdwarfs, for which it has already proven effective. By
definition, M subdwarfs are stars with $\zeta<$0.82
\citep{Lepine2007}. Though we have a few stars with values of $\zeta$
just marginally under 0.82, only one star clearly stands out as a
definite M subdwarf: the star PM I20050+5426 (= Gl 781) which boasts a
$\zeta=0.58$ well within the M subdwarf regime. The star also clearly
stands out in Figure~\ref{cah_tio} where it lies noticeably below the
main locus at TiO5$_{c}\simeq$0.75.

PM I20050+5426 is also known as V1513 Cyg, a star previously
identified as an M subdwarf by \citet{Gizis1997}, and one clearly
associated with the Galactic halo \citep{Fuchs1998}. The star is also
notorious for being a flare star, with chromospheric
activity due not to young age but to the presence of a low-mass
companion on a close orbit \citep{Gizis1998}. Our own spectrum indeed
shows a relatively strong line of H$\alpha$ in emission, which is
extremely unusual for an M subdwarf. It is an interesting coincidence
that the brightest M subdwarf in the northern sky should turn out to
be a peculiar object.

In any case, because the TiO molecular bands are weaker in M subdwarfs
than they are in M dwarfs, the use of TiO spectral indices for
spectral classification leads to underestimates of their spectral
subtype. The convention for M subdwarfs is rather to base the
classification on the strengths of the CaH bandheads
\citep{Gizis1997,Lepine2003,Lepine2007}. We adopt the same convention
here, and recalculate the subtype from the mean of Equations 6 and 7
only (CaH2 and CaH3 indices). We thus classify PM I20050+5426 as an
sdM2.0, which is one half-subtype later than the sdM1.5 classification
suggested by \citep{Gizis1997}.

\subsection{Photometric dependence on metallicity}

A prediction of current atmospheric models is that metallicity
variations in M dwarfs yield significant variations in optical
broadband colors \citep{Allard2000}. The metal-poor M subdwarfs have
in fact long been known to have bluer V-I colors than the more
metal-rich field M dwarfs of the same luminosity
\citep{Monet_etal.1992,Lepine2003}. The bluer colors are due to
reduced TiO opacities in the optical, which make the spectral energy
distribution of M subdwarfs closer to that of a blackbody, while it
makes the metal-rich M dwarfs display extreme red colors.

Interestingly, the SDSS $g-r$ color index shows the opposite trend,
and is bluer in the more metal-poor stars. This is because the TiO
bands very strongly depress the flux in the 6000\AA-7000\AA\ (r-band)
range, an effect which in fact makes the metal-rich M dwarfs
degenerate in $g-r$, as the increased TiO opacities in cooler stars
balance out the reduced flux in $g$ from lower T$_{eff}$. This effect
is much weaker in metal-poor stars due to the reduced TiO opacity,
which makes metal-poor stars go redder as they are cooler, as one
would normally expect. This has been observed in late-type M
subdwarfs, which have significantly redder color that field M dwarfs
\citep{LepineScholz2008}. The color dependence of M dwarfs/subdwarfs
on metallicity is also predicted by atmospheric
models. Figure~\ref{model} (bottom-right panel) shows the predicted
$g-r$ and $r-z$ colors from the PHOENIX/BT-SETTL model of
\citet{Allard}. The models corroborate observations and predict redder
$g-r$ colors in metal-poor stars.

Although $ugriz$ photometry is not available for our stars (all of them
are too bright and saturated in the Sloan Digital Sky Survey), it is
possible to use the well-calibrated SNIFS spectrophotometry to
calculate synthetic broadband $riz$ magnitudes for the subset of stars
observed at UH. We first examine any possible correlation between the
optical to infrared $r-K_S$ color (taken as a proxy for V-I) and the
$\zeta$ index. Figure~\ref{model} plots $r-K_s$ as a function of
spectral subtype, with the dots color-coded for the $\zeta$ values of
their associated M dwarf (top-left panel). We find a tight
relationship between $r-K_s$ and spectral subtype, which we fit using
a running median. The residuals are plotted in the top-right panel,
and show no evidence of a correlation with $\zeta$. There are a
significant number of outliers with redder $r-K_s$ colors than the
bulk of the M dwarfs: these likely indicate systematic errors in
estimating the synthetic $r$ band magnitudes. The absence of any clear
correlation suggests that an optical-to-infrared color such as $r-K_S$
is not sensitive enough to detect small metallicity variations, at
least at the metal-rich end.

The synthetic $g-r$ and $r-z$ colors are plotted in Figure~\ref{model}
(lower-left panel). The redder stars ($r-z$>1.2) show a wide scatter
in $g-r$, on the order of what is predicted for stars with a range of
metallicities $-0.5<[Fe/H]<0.5$. Though we do not find a clear trend
between the synthetic $g-r$ colors and the $\zeta$ values measured in
the same stars, the high-$\zeta$ stars (red and orange dots on the
plot) do seem to have lower values of $g-r$ on average than the
low-$\zeta$ ones (green dots). The trend is suggestive of a
metallicity link to both the $g-r$ colors and the $\zeta$ values, and
should be investigated further with data of higher precision.

%
%
%
%

\section{Chromospheric activity}

\begin{deluxetable*}{lrrrrrrcccc}
\tabletypesize{\scriptsize}
\tablecolumns{11} 
\tablewidth{0pt} 
\tablecaption{Survey stars: distances, kinematics, and activity.\label{table_distance}}
\tablehead{
\colhead{Star name} & 
\colhead{$\pi_{trig}$} &
\colhead{$\pi_{phot}$} &
\colhead{$\pi_{spec}$} &
\colhead{U} &
\colhead{V} &
\colhead{W} &
\colhead{EWHA} &
\colhead{H$\alpha$} &
\colhead{Xray} &
\colhead{UV} \\
\colhead{} &
\colhead{$\arcsec$} &
\colhead{$\arcsec$} &
\colhead{$\arcsec$} &
\colhead{km s$^{-1}$} &
\colhead{km s$^{-1}$} &
\colhead{km s$^{-1}$} &
\colhead{\AA} &
\colhead{active} &
\colhead{active} &
\colhead{active}
}
\startdata  
PM I00006+1829  &       \nodata      &    \nodata        &    \nodata        &\nodata&\nodata&\nodata&\nodata& -& -& -\\
PM I00012+1358S &       \nodata      &  0.030$\pm$  0.008&  0.028$\pm$  0.008&  -13.9&   12.1&\nodata&   0.41& -& -& -\\
PM I00033+0441  &  0.0342$\pm$ 0.0032&  0.031$\pm$  0.008&  0.030$\pm$  0.009&    8.5&   -6.8&\nodata&   0.39& -& -& -\\
PM I00051+4547  &  0.0889$\pm$ 0.0014&  0.078$\pm$  0.021&  0.083$\pm$  0.024&  -38.2&\nodata&  -15.9&   0.47& -& -& -\\
PM I00051+7406  &       \nodata      &    \nodata        &    \nodata        &\nodata&\nodata&\nodata&\nodata& -& -& -\\
PM I00077+6022  &       \nodata      &  0.078$\pm$  0.031&  0.091$\pm$  0.027&  -15.1&\nodata&   -4.4&  -3.11& Y& -& -\\
PM I00078+6736  &       \nodata      &  0.046$\pm$  0.012&  0.055$\pm$  0.016&    5.1&\nodata&   -8.5&   0.39& -& -& -\\
PM I00081+4757  &       \nodata      &  0.071$\pm$  0.028&  0.061$\pm$  0.018&    7.9&\nodata&    1.8&  -2.93& Y& -& Y\\
PM I00084+1725  &  0.0460$\pm$ 0.0019&  0.040$\pm$  0.011&  0.040$\pm$  0.012&   11.2&    0.7&\nodata&   0.41& -& -& -\\
PM I00088+2050  &       \nodata      &  0.078$\pm$  0.031&  0.080$\pm$  0.024&    9.5&\nodata&  -10.1&  -4.64& Y& -& Y\\
PM I00110+0512  &  0.0233$\pm$ 0.0038&  0.032$\pm$  0.009&  0.031$\pm$  0.009&  -48.5&  -14.3&\nodata&   0.35& -& -& -\\
PM I00113+5837  &       \nodata      &    \nodata        &    \nodata        &\nodata&\nodata&\nodata&\nodata& -& -& -\\
PM I00118+2259  &       \nodata      &  0.055$\pm$  0.022&  0.054$\pm$  0.016&   -3.0&\nodata&  -16.7&   0.30& -& Y& -\\
PM I00125+2142En&  0.0358$\pm$ 0.0028&  0.023$\pm$  0.006&  0.024$\pm$  0.007&   -5.9&\nodata&  -32.5&   0.44& -& -& -\\
PM I00131+7023  &       \nodata      &  0.037$\pm$  0.010&  0.037$\pm$  0.011&   -6.2&\nodata&   16.8&   0.37& -& -& -
\enddata
\end{deluxetable*} 

\begin{figure}
\vspace{-0.3cm}
\hspace{-0.6cm}
\includegraphics[scale=0.60]{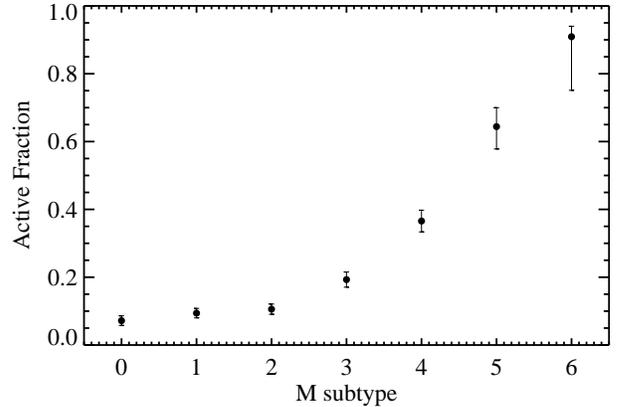}
\caption{Fraction of active stars as a function of the spectral
  subtype M. The rise at later subtypes is consistent with earlier
  studies of field M dwarf, which shows increased activity levels in
  mid-type M dwarfs. The fraction level at later subtype is however
  higher than that measured in the SDSS spectroscopic
  catalog.\label{frac_active}}
\end{figure}

%
%
%

To evaluate the presence of H$\alpha$ in emission in our M dwarfs, we
used the H$\alpha$ equivalent width index EWHA defined as:
\begin{equation}
EWHA = 100{\rm \AA} \left[ 1 - \frac{ 14{\rm \AA}
    \int_{6557.61}^{6571.61} S(\lambda)
    d\lambda}{ 100{\rm \AA} \left( \int_{6500}^{6550} S(\lambda) d\lambda +
  \int_{6575}^{6625} S(\lambda) d\lambda \right)} \right],
\end{equation}
where $S(\lambda)$ is the observed spectrum. The EWHA index measures
the flux in a region (6557.61\AA-6571.61\AA), which includes the
$H\alpha$ line, in relation to a pseudo-continuum region spanning
6500\AA-6550\AA\ and 6575\AA-6625\AA; the calculation provides a value
in units of wavelength (\AA) like the traditional equivalent
width. Note that for an $H\alpha$ line in emission, values of the EWHA
index are negative, following convention. Assuming the W1-W2 region to
measure the true spectral continuum, then the EWHA index would measure
the true equivalent width of $H\alpha$. As it turns out, the $W1-W2$
region often has a higher mean flux than the $W3-W4$ region without
the $H\alpha$ emission component, which means that the EWHA index
systematically underestimate the equivalent width of the $H\alpha$
line. The index is however reproducible and more convenient for
automated measurement than, e.g. manual evaluation of the equivalent
using interactive software such as IRAF.

The EWHA index was measured for all spectra in our sample, and used to
flag active stars. Following \citet{West2011}, we defined a
star to be chromospherically ``active'' if EWHA $<-0.75\AA$, which
usually corresponds to a clearly detectable $H\alpha$ line in
emission. Values of the EWHA index are listed in
Table~\ref{table_distance}. Under the above criterion, 171 M dwarfs in
our survey are considered active.

\citet{Hawley1996} found that active stars (by their criterion,
EW(H$\alpha$) $> 1 \AA$) have slightly redder $V-K$ colors for the
same value of TiO5 index, an effective temperature proxy.  We
calculated a median ($N=30$) locus of $V-K$ vs. TiO5 index for our
entire sample and found that 13 active stars are bluer than this
locus, while 45 are redder, seemingly confirming their result. The
large scatter in $V-K$ colors however prevents us from quantifying
this offset more precisely.

The fraction of stars that are active at each spectral subtype is shown
in Figure~\ref{frac_active}, with error bars computed from the binomial
distribution. The increase in the active fraction with spectral type
is consistent with previous studies
\citep{JoyAbt1974,Hawley1996,West2004,West2008,Kruse_etal.2010,West2011}. 
Our active fractions are higher at subtype M4-M6 than the
\citet{Hawley1996} results, even when using their criterion to define active
stars (see above). This may be a result of a slightly different
definition for EW, or a result of the relatively small number of
objects. Our active fractions are also higher than the \citet{West2011}
results at each subtype. This discrepancy is likely caused by the
magnitude limit imposed in our survey: our objects are all nearby, and
relatively close to the Galactic plane (see Section 8), which makes
them statistically younger, as also suggested \citet{West2011}. The
active fractions for our stars are closer to the active fractions for
the \citet{West2011} stars in bins of stars closest to the Galactic
plane.

Another chromospheric activity diagnostic in M dwarfs is the detection
of X-rays. M dwarfs that are X-ray bright are often young, and this has
been used to identify members of nearby young moving groups
\citep[e.g.,][]{Gaidos1998, Zuckerman2001, Torres2006}, and other
young stars in the Solar Neighborhood \citep{Riaz2006}. Most recently,
\citet{Shkolnik2009,Shkolnik2012} and \citet{Schlieder2012} have used
the ratio of {\it ROSAT} X-ray flux to 2MASS {\it J} or {\it K}-band
flux to identify candidate members of young moving groups. This
technique is particularly effective for objects $\lesssim70$pc away,
which includes all the stars in our sample. The
\citet{LepineGaidos.2011} catalog, from which our targets are drawn,
was already cross-matched to the {\it ROSAT} All-Sky Bright Source
Catalog \citep{Voges.etal.1999} and the {\it ROSAT} All-Sky Survey
Faint Source Catalog \citep{Voges.etal.2000}.

\begin{figure}
\vspace{-0.3cm}
\hspace{-0.6cm}
\includegraphics[scale=0.53]{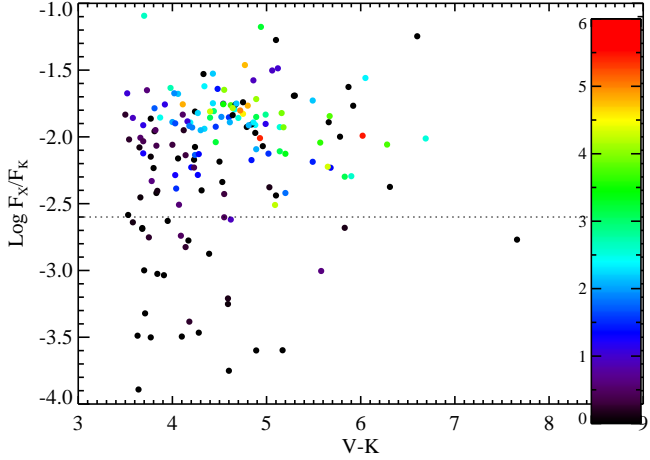}
\caption{X-ray luminosity normalized by the flux in the infrared K$_s$
band, plotted as a function of the optical-to-infrared color $V-K$,
for stars in our sample which have counterparts in the ROSAT all-sky
points source catalog. The color scheme shows the H$\alpha$ equivalent
width; active stars are found to have large X-ray flux, as expected
from chromospheric activity.\label{act_xray}}
\end{figure}

We have computed the X-ray flux for our survey stars from the measured
count rate and hardness ratio (HR1) using the prescription in
\citet{Schmitt1995}. Figure~\ref{act_xray} shows the distribution of
X-ray flux as a function of $V-K$ color, for the 290 M dwarfs with
{\it ROSAT} detections. Dots are color-coded according to the strength
of the H$\alpha$ emission, as measured by the EWHA index. As
expected, M dwarfs with strong H$\alpha$ emission also tend to be more
X-ray bright. Objects with $\log F_X \ F_K > -2.6$ (above the dashed
line in Figure~\ref{act_xray}) are considered bright enough in X-ray
to qualify as chromospherically active, following the definition of
\citet{Schlieder2012}. Some 154 of the X-ray sources are active
based on their EWHA values, and all of them also qualify as active
stars based on their X-ray fluxes. On the other hand, 53 M dwarfs
identified as active based on X-ray flux do not display significant
H$\alpha$ emission in our spectra; most of them tend to be earlier M
dwarfs, in which H$\alpha$ emission is not as easily detected as in
later type objects because of their higher continuum flux near
$\lambda6563\AA$. There are also 22 stars in our survey which are active
based on $H\alpha$ but are not detected by ROSAT. This suggests that
only two thirds of the ``active'' stars will be diagnosed as such from
both X-ray and H$\alpha$ emission, while the other third will show
only either. This could be due to source confusion in the ROSAT X-ray
survey, variability in either X-ray or H$\alpha$ emission, or, in the
case of the X-ray flux, non-uniform sky coverage by ROSAT. We
calculated the luminosity ratio index $L_X/L_{H\alpha}$ of active
stars, as defined by the criterion EW(H$\alpha$) $>1 \AA$ following
the procedure of \citet{Hawley1996}, and adopting the relation $V-R
\approx 0.7 + 0.06 {\rm SpTy}$ to estimate an $R$ magnitude and the
continuum flux at the H$\alpha$ line. We find that the ratio is
insensitive to bolometric magnitude and spectral type, and has a
median value of 0.85. This is higher than the \citet{Hawley1996}
average of $\sim 0.5$, and may in part be due to Malmquist bias in the
flux-limited ROSAT survey favoring the inclusion of the most X-ray
luminous stars, as well as greater variation in the ratio because of
the elapsed time (two decades) between the ROSAT survey and
our observations.

\begin{figure}
\vspace{-0.3cm}
\hspace{-0.6cm}
\includegraphics[scale=0.53]{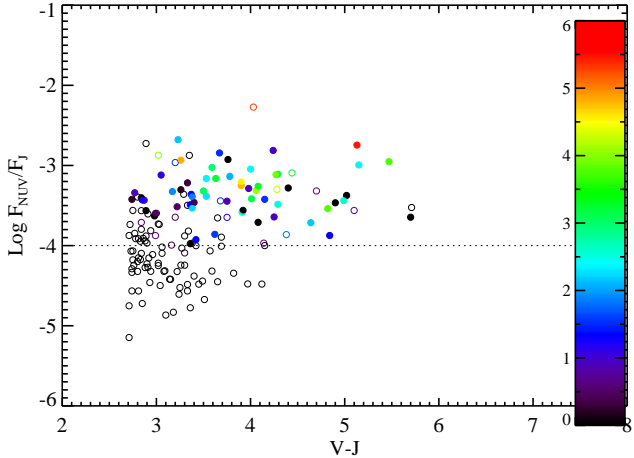}
\caption{Normalized near-UV flux as a function of the
  optical-to-infrared $V-K_s$ color. The color scheme shows the
  strength of the H$\alpha$ equivalent width. Closed circles show
  stars identified as active based on X-ray emission, closed circles
  show stars with low or no detection in ROSAT. \label{act_uv}}
\end{figure}

Active stars can also be identified from ultra-violet
excess, as suggested in Section 3.4. \citet{Shkolnik2011,Shkolnik2012}
showed that {\it GALEX} UV fluxes can identify young M dwarfs in
nearby moving groups, and can identify active stars to larger
distances. Figure~\ref{act_uv} shows the {\it GALEX} NUV to 2MASS
{\it J} flux ratio, for the objects with UV detections. The dashed
line shows the selection criteria of \citet{Shkolnik2011}. Dot colors
represent the EWHA index values for the stars, while filled circles
indicate objects with $\log F_X / F_K > -2.6$, i.e. stars whose X-ray
flux does not identify them as being active. Overall, there is a good
correpondence between the different activity diagnostics. However,
there are some stars identified as active based on UV flux that are
not identified as such from their X-ray and/or $H\alpha$
emission. Again this suggests that a complete identification of active
M dwarfs in the solar vicinity may require a combination of diagnostic
features.

In any case our survey, which combines X-ray, UV, and $H\alpha$
diagnostics, provides a valuable subset for identifying low-mass young
stars in the Solar Neighborhood, and may potentially yield new members
of young moving groups, or even the identification of new moving
groups. The last three columns in Table~\ref{table_distance} display
flags for stars found to be active from either H$\alpha$, X-ray, or UV
flux. The flag indicates activity by a ``Y''. Absence of a flag does not
necessarily indicate absence of activity: the GALEX survey does not
cover the entire sky, and the ROSAT X-ray survey is not uniform in
sensitivity, so a non-detection in either does not necessarily
indicate quiescence. Activity diagnostics could also be
time-variable. H$\alpha$ equivalent width is particular are know to be
variable on various timescales \citep{Bell2012}. In any case, there is
a good correlation between the different diagnostics. We flag 175
stars as active based on H$\alpha$, 42 based on X-ray emission, and
172 based on UV excess. Overall, 252 stars are assigned one or more
activity flags: 19 stars have all three flags on, 99 stars get two
flags, and 137 get only one.

\section{Distances and kinematics}

\subsection{Spectroscopic distances}
\label{sec:distances}

\begin{figure}
\vspace{0.0cm}
\hspace{-0.2cm}
\includegraphics[scale=0.85]{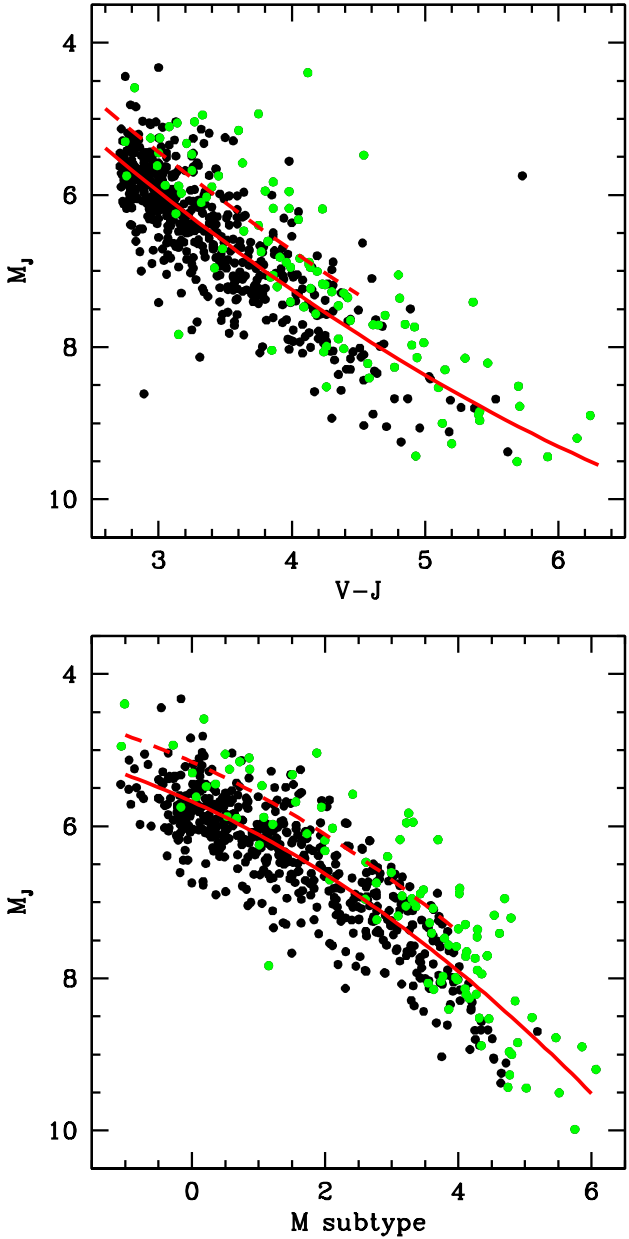}
\caption{Absolute visual ($M_V$) and infrared ($M_J$) magnitudes for
  the 631 stars in our survey for which geometric parallax
  measurements exist in the literature. Top panels: absolute
  magnitudes against $V-J$ color, which follow the color-magnitude
  relationship used in \citet{LepineGaidos.2011} to estimate
  photometric distances. Bottom panel: absolute magnitudes as a
  function of spectral subtype, based on the spectral-index
  classification described in this paper. Active stars are plotted
  in green, and are found to be overluminous at a given $V-J$ color
  and given spectral subtype, compared with non-active stars. The
  offset in notable for stars of earlier M subtypes (or bluer $V-J$
  colors).\label{dist_cal}}
\end{figure}

Astrometric parallaxes are available for 631 of the stars in our
sample, spanning the full range of colors and spectral subtypes. We
combine these data with our spectroscopic measurements to re-evaluate
photometric and spectroscopic distances calibrations for M dwarfs in
our census. Absolute visual magnitudes $M_V$ are calculated and are
plotted against both $V-J$ color and spectral subtype in
Figure~\ref{dist_cal}. M dwarfs with signs of activity (H$\alpha$, UV,
X-ray) are plotted in green, other stars are plotted in
black. The solid red lines are the best-fit second order polynomials,
when both active and inactive stars are used, and after elimination of
3$\sigma$ outliers. The equations for the fits, where
$spT$ is the spectral type (K7 is -1 and M0=0, etc) are: 
\begin{eqnarray}
M_J = 1.194 + 1.823 (V-J) - 0.079 (V-J)^2\\
M_J = 5.680 + 0.393 (SpT) + 0.040 (SpT)^2
\label{eqn:mj_vs_spt}
\end{eqnarray}
where SpT are the spectral types, and with the least-squares fit
performed after exclusion of 3-sigma outliers. The 1$\sigma$
dispersion about these relationships are $\pm0.61$ mag for
(M$_J$,V-J), and $\pm0.52$ mag for (M$_J$,SpT). The smaller scatter in
the spectroscopic relationship suggests that spectroscopic distances
may be marginally more accurate than the photometric ones. We suspect
that the larger uncertainty on the photographic $V$ magnitudes may be
the cause.

The most notable feature in the diagrams is that active stars appear
to be systematically more luminous than non-active stars. This
corroborates the observation made previously by \citet{Hawley2002}, as
part of the PMSU survey. \citet{Hawley2002} found that active stars
were more luminous by 0.48mag in a diagram of $M_K$ against TiO5
index, used as proxy for spectral subtype. For stars in our census, we
find that among stars of spectral subtype M2 and earlier, active stars
are on average 0.46mag more luminous at a given subtype than
non-active stars; in the color-magnitude diagram, bluer stars of
colors $V-J<4.0$ which are active are on average 0.47mag more luminous
than non-active stars of the same color. Both values agree well with
the values quoted by \citet{Hawley2002}. The systematic overluminosity
of active stars is also responsible for some of the scatter in the
color-magnitude and spectral-type magnitude relationships. If we
exclude active stars, the scatter about the color-magnitude
relationship fals marginally to $\pm0.58$, and the scatter in the
spectral-type magnitude relationship falls to $\pm0.49$. 

The offset in absolute magnitude between active and non-active stars
suggests that spectroscopic and photometric distances would be more
accurate for active stars in our census if their estimated absolute
magnitudes were made 0.46mag brighter that suggested by
Equation~\ref{eqn:mj_vs_spt}. We therefore adopt the following
relationships to be applied only on active stars of subtype M2.5 and
earlier:
\begin{eqnarray}
{[M_J]}_{early-active} = 0.734 + 1.823 (V-J) - 0.079 (V-J)^2\\
{[M_J]}_{early-active} = 5.220 + 0.393 (SpT) + 0.040 (SpT)^2 
\label{eqn:mj_vs_spt2}
\end{eqnarray}
Again we define as ``active'' any star which qualifies as such base on
any one of our criteria (H$\alpha$, UV, X-ray). There are several
reasons that would explain why active, early-type stars are more
luminous at a given subtype. Activity in an early-type M dwarf could
mean that the star is younger \citep{Delfosse_1998}; early-type M
dwarfs with ages $<100$Myr are known to be overluminous at a given
color, due to lower surface gravity \citep{Shkolnik2012}. Older stars
could remain active due to interaction with a close companion
\citep{Morgan_2012}, in which case the active stars would also appear
overluminous due to this unresolved companion. Late-type M dwarfs,
however, can remain active for long periods of time, and would thus
not require the star to be young or have a close companion.

Splitting the active and inactive stars results in lowering the
scatter of non-active stars in the subtype-magnitude
relationship (to $\pm$0.5mag). Overall, our spectroscopic distances
for non-active M dwarfs provide a $1\sigma$ uncertainty of $\pm26\%$
on the distance. For active stars, we find a scatter of $\pm$0.6mag,
which suggests distance uncertainties of $\pm32\%$. Our photometric
distances estimated from (V,V-J) have similar though perhaps slightly
larger uncertainties. Photometric and spectroscopic parallaxes,
estimated from the above relationships for active and non-active
stars, are listed in Table~\ref{table_distance}.

We estimated the effect of two relevant sampling biases on the
calibration between $M_J$ and $V-J$ color. In Eddington bias,
photometric errors scatter more numerous, bluer, and intrinsically
brighter stars to redder apparent colors than redder stars are
scattered to bluer apparent colors \citep{Eddington1913}. The net
effect is to make stars at a given apparent color appear more luminous
than they are. In Lutz-Kelker (LK) bias, a form of Malmquist bias,
errors in trigonometric prallax will scatter more numerous and more
distant stars with lower parallax to higher apparent parallax values,
making them appear less luminous than they are \citep{Lutz1973}. By
taking the derivative of $M_J$ with respect $V-J$ color, multiplying
by the derivative of the number of stars in our $J$-magnitude-limited
catalog with respect to $M_J$, assuming that the errors in $V-J$ are
gaussian-distributed with standard deviation $\sigma_{V-J}$, and
integrating over the distribution, we find the Eddington bias in $M_J$
to be;
\begin{equation}
\Delta_E = -\ln 10 \left[1.918 = 0.178\left(V-J\right)\right]^2 \left(0.6 -
0.4\gamma\right) \sigma_{V-J}^2,
\end{equation}
where $\gamma$ is the power-law index of a luminosity function for M
stars which we take to be 0.325. Performing a similar derivation for
the effect of L-K bias on $M_J$, we find:
\begin{equation}
\Delta_{LK} = \frac{15}{\ln 10} \sigma_{\pi}^2,
\end{equation}
where $\sigma_{\pi}$ is the fractional error in parallax. Using
published parallax errors for our {\it Hipparcos} stars and adopting a
conservative $\sigma_{V-J} = 0.05$, we find that L-K bias usually
dominates over Eddington bias and that 74\% of our stars have a total
bias of less than +0.2 magnitudes. A running median ($N = 100$)
vs. $V-J$ color is highest ($\sim 0.15$) for the bluest ($V-J =
2.7$) stars and falling to less than +0.05 magnitudes for $V-J > 3.3$.
An analogous analysis can be performed for the bias in $M_J$
vs. spectral type, with a similar result. To debias values of $M_J$,
these values should be {\it subtracted} from our calibration but we do
not perform that operation here because of the small magnitude of the
effect, which would overestimate distances by about 2\% on
average. The correction would also seem negligible compared with the
intrinsic scatter in our color-magnitude and subtype-magnitude
relationships are of order $\pm0.5-0.6$, much larger than the L-K
correction. 

\begin{figure}
\vspace{-0.3cm}
\hspace{-0.2cm}
\includegraphics[scale=0.44]{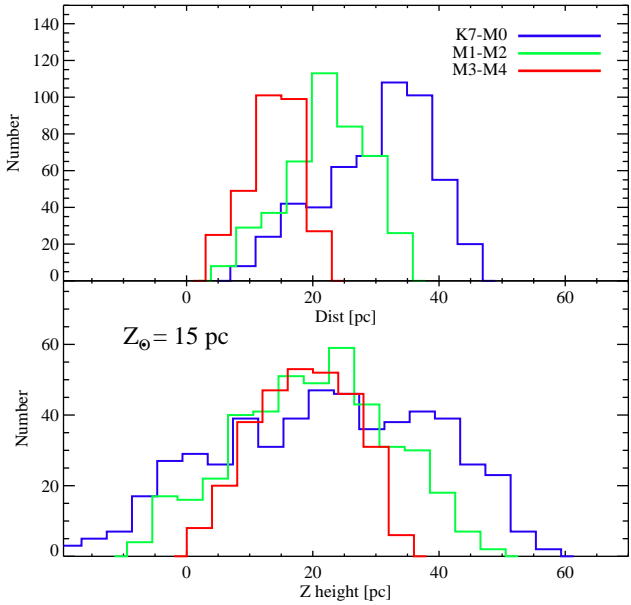}
\caption{Top: distribution of spectroscopic distances for the stars in
  our survey, shown for three ranges of spectral subtypes. Early-type
  stars are clearly sampled over a larger volume, which explains why
  they dominate our survey. Bottom: distribution of Galactic scales
  heights for the same stars, assuming that the Sun is hovering 15~pc
  sbove the Galactic midplane. As expected from our magnitude-limited
  sample, stars of later spectral subtypes (and lower luminosity) are
  found at shorter distances. Our survey samples a region well within
  the Galactic thin disk.\label{disthist}}
\end{figure}

Figure \ref{disthist} shows the distribution of photometric distances
for our complete sample using Equation \ref{eqn:mj_vs_spt} and the
M$_J$=M$_J$(V-J) color-magnitude relationship. The spectral subtypes are
plotted in separate colors and demonstrates that the earlier-type
stars, which are intrinsically brighter, are sampled to significantly
larger distances compared with the later-type stars. In the 20pc
volume, the M3-M4 stars still appear to dominate. We also plot the
Galactic height of the stars in our sample, adopting a Galactic height
of 15 pc for the Sun \citep{Cohen1995,Ng1997,Binney1997}. It is clear
that our survey is largely contained within the Galactic thin disk,
and barely extends south of the midplane. This is consistent with the
relative absence of metal-poor stars associated with the thick disk
and halo.

\subsection{Kinematic analysis}

\begin{figure*}
\vspace{-0.3cm}
\hspace{0.5cm}
\includegraphics[scale=0.82]{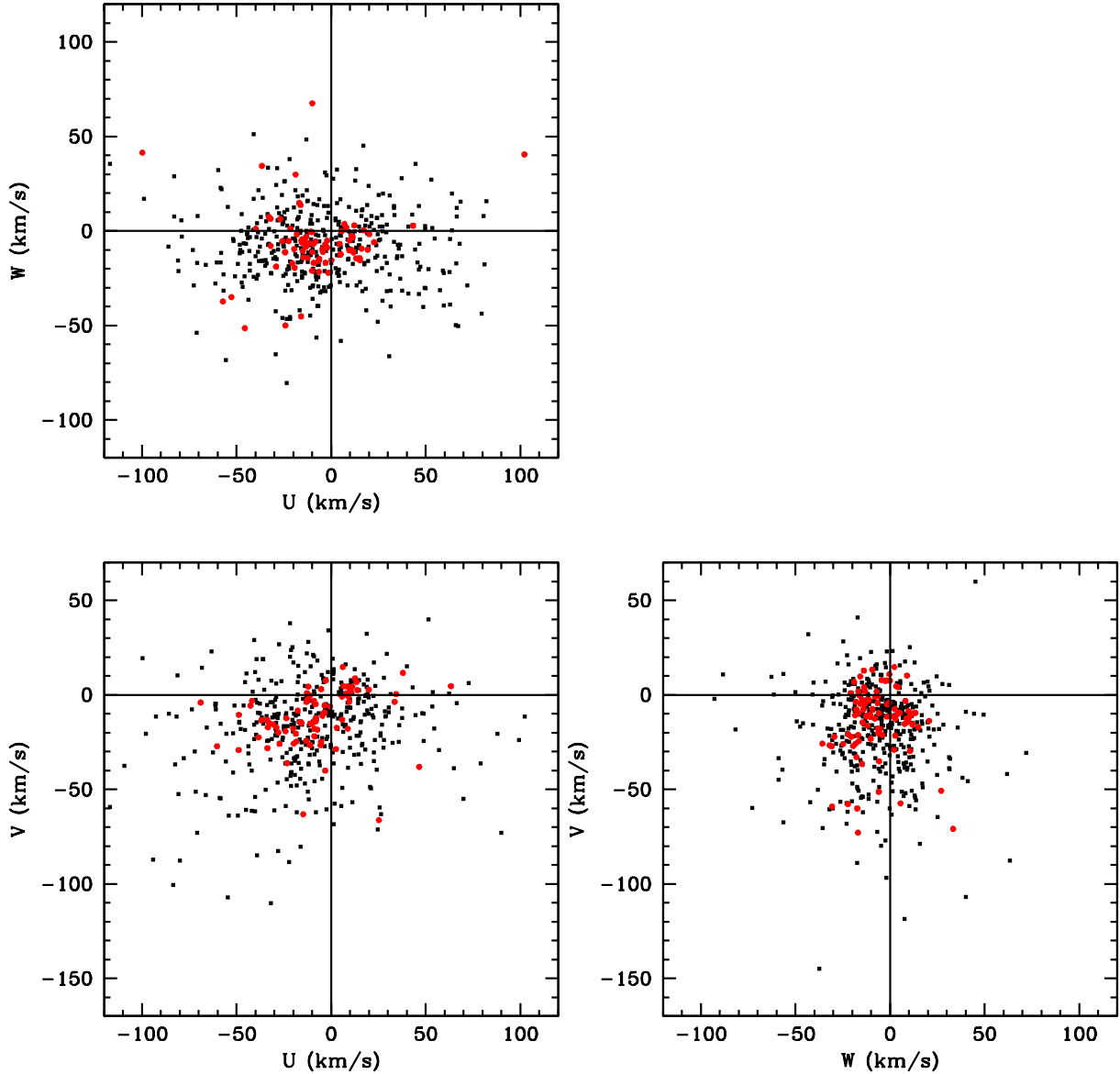}
\caption{Velocity-space projections for the M dwarfs in our
  survey. Velocies are calculated based on photometric distances and
  proper motions alone (no radial velocities used). Each star in our
  census is thus displayed in only one panel, which corresponds to the
  projection in which the radial velocity of the star has the smallest
  contribution. Stars with significant levels of H$\alpha$ emission,
  i.e. chromospherically active M dwarfs, are plotted in
  red. \label{uvw}}
\end{figure*}

Accurate radial velocities are not available for most of the stars in
our sample, which prevents us from calculating the full (U,V,W)
components of motion for each individual star. However, it is possible
to use the distance measurements (for stars with parallaxes) or
estimates (for stars with no parallax), and combine them to the proper
motions to evaluate with some accuracy at least two of these
components for each star. More specifically, we calculate the (U,V,W)
components by assuming that the radial velocities $R_V=0$. We then
consider the (X,Y,Z) positions of the stars in the Galactic reference
frame, from the distances and sky coordinates. For stars with the
largest component of position in +X or -X, the radial velocity mostly
contibute to U, and has minimal influence on the values of V and
W. Likewise stars with the largest component of position in +Y or -Y
(+Z or -Z) make good tracers of the velocity distribution in U and W
(U and V). We use this to assign any one of (U,V) or (U,W) or (V,W)
velocity component doublet to every M dwarf in our catalog.
Estimated values of the components of velocity are listed in
Table~\ref{table_distance}. For each star, one of the components is
missing, which is the component that would most depend on the radial
velocity component based on the coordinates of the star. Again, the
other two components are estimated only from proper motion and
distance. For the distance, we use the trigonometric parallaxes
whenever available; otherwise the spectroscopic distances as are used,
based on the raltionships described in the previous section.

The resulting velocity distributions are displayed in
Figure~\ref{uvw}. We measure mean values of the velocity components
using all allowable values and find:
\begin{displaymath}
 <U> =  -8.1~{\rm km s^{-1}}, \sigma_U = 32.8~{\rm km s^{-1}},
\end{displaymath}
\begin{displaymath}
 <V> = -17.0~{\rm km s^{-1}}, \sigma_V = 22.8~{\rm km s^{-1}},
\end{displaymath}
\begin{displaymath}
 <W> =  -6.9~{\rm km s^{-1}}, \sigma_W = 19.3~{\rm km s^{-1}}.
\end{displaymath}
The values are also largely consistent with those found from the PMSU
survey and desribed in \citep{Hawley1996}. They are also remarkably
similar to the moments of the velocity components calculated by
\citet{Fuchs2009} for SDSS stars of the Galactic thin disk, and which
are $<U> = -8.6, \sigma_U = 32.4$, $<V> = -20.0, \sigma_V = 23.0$, and
$<W> =  -7.1, \sigma_W = 18.1$. The agreement suggests that our
distance estimates are reasonably accurate, and it corroborates
earlier results about the kinematics of the local M dwarf population
which indicate a larger scatter of velocities in $U$. The mean values
of $<U>$ and $<V>$ are consistent with the offsets from the local
standard of rest as described in \citet{DehnenBinney_1998}

Active stars are found to have significantly smaller dispersions in
velocity space. All 252 stars with at least one activity flags
(i.e. stars found to be active either from H$\alpha$, X-ray flux, or
UV excess) are plotted in red in Figure~\ref{uvw}. Those active stars
have first and second moments:
\begin{displaymath}
 <U> =  -9.3~{\rm km s^{-1}}, \sigma_U = 25.2~{\rm km s^{-1}},
\end{displaymath}
\begin{displaymath}
 <V> =  -13.4~{\rm km s^{-1}}, \sigma_V = 16.8~{\rm km s^{-1}},
\end{displaymath}
\begin{displaymath}
 <W> =  -6.6~{\rm km s^{-1}}, \sigma_W = 15.3~{\rm km s^{-1}}.
\end{displaymath}
The values are consistent with \citep{Hawley1996}, who also reported
that active M dwarfs tend to have a smaller scatter compared with
inactive M dwarfs. The smaller dispersion values suggest that these
active stars may be significantly younger than the average star in the
Solar Neighborhood.

In any case, we also note that the velocioty space distribution is
non-uniform and shows evidence for substructure. Our M dwarf data
shows velocity-space substructure as that observed and described in
\citet{Nordstrom2004,Holberg2008} for solar-type stars in the vicinity
of the Sun. This substructure is sometimes referred to as ``streams''
or ``moving groups'', although an analysis by \citet{Bovy_Hog2010}
shows that these groups do not represent coeval populations arising
from star-formation episodes. The velocity-space substructure is more
likely transient and associated with gravitational perturbations which
are the signature of the Galactic spiral arms
\citet{QuillenMinchev2005} and the Galactic bar \citet{Minchev2012}. A
simple description of the velocity-space distribution in terms of
means values and dispersions, or as velocity ellipsoid, is therefore only a
crude approximation of a more complex and structured distribution.

Finally, we note that the stars of our catalog that were previously
part of the CNS3 and stars with measured trigonometric parallaxes
(e.g. from the Hipparcos catalog) tend to have larger velocity
dispersions, with $(\sigma_U,\sigma_V,\sigma_W)$=(38.4,25.9,23.1),
while the newer stars have
$(\sigma_U,\sigma_V,\sigma_W)$=(26.2,19.4,15.3). The difference could
be due to systematic underestimation of the photometric/spectroscopic
distances, but a more likely explanation is that the CNS3 and parallax
subsample suffers from proper motion selection. This is because most
of the CNS3 stars and M dwarfs monitored with Hipparcos were selected
from historic catalogs of high proper motion stars, which have a
higher limit than the SUPERBLINK proper motion catalog used in the
LG2011 selection. This kinematic bias means that the current subset of
M dwarfs monitored for exoplanet programs suffers from the same
kinematic bias, which could possibly introduce age and metallicity
selection effects.

\section{Conclusions}

We have now compiled spectroscopic data for a nearly complete list of
M dwarfs in the northern sky with apparent magnitudes $J<9$. Our
survey identifies a total of 1,403 very bright M dwarfs. Our new
catalog provides spectral subtypes and activity measurements
($H\alpha$ emission) for all stars, as well as a rough indicator of
metallicity in the guise of the $\zeta$ parameter, which measures the
ratio of TiO to CaH bandstrengths. Only one of the stars in the survey
is unambiguously identified as a metal-poor M subdwarf (PM I20050+5426
= V1513 Cyg).

Our target stars were identified from the all-sky catalog of bright M
dwarfs presented in \citet{LepineGaidos.2011}. As such, our
spectroscopic survey suffers from the same selection effects and
completeness issues. The completeness and bias of the SUPERBLINK
proper motion survey, from which these stars were selected is
discussed a length in \citet{LepineGaidos.2011}. In the northern
hemisphere, the SUPERBLINK catalog is complete for proper motions $\mu
>40$~mas$^{-1}$. We show in \S4 that there is a kinematic bias in the
catalog which excludes stars with very low transverse motions (in the
plane of the sky), but the low proper motion limit means that less
than 5\% of stars within 65 parsecs of the Sun are in fact excluded in
the selection. In addition, we estimate that $\approx$5\% of the nearby,
bright M dwarfs may have escaped our target selection scheme due to
faulty magnitudes. Therefore, we estimate that our census most likely
include $>90\%$ of all existing M dwarfs in the northern sky with
$J<9$. Early-type K7-M1 dwarfs have absolute magnitudes
$M_J\approx5.5$, and our $J<9$ sample thus identifies them well to a
distance of about 50pc, as confirmed in Figure~\ref{disthist}. On the
other hand, later type M3-M4 dwarfs have $M_J\approx8$ and thus only
those at very close distance range ($<$15~pc) will be included in the
catalog. Their completeness will however be very high because the
proper motion bias excludes less than $1\%$ of the stars within that
distance range. In any case, the different survey volumes for
early-type and late-type stars means that our survey favors the former
over the latter by a factor of about 35 to 1. It is thus no surprise
that our spectroscopic catalog is dominated by early-type M dwarfs.

An important result of our spectroscopic analysis is the
identification of systematic errors in the spectral indices, which
measure the strenghts of the CaH, TiO, and VO molecular
bands. Systematic offsets between data obtained at MDM Observatory and
at the University of Hawaii 2.2-meter telescopes, as well as offsets
between these and the values measured for the sames stars in the
Palomar-MSU survey of \citet{Reid1995}, indicate that these spectral
indices are susceptible to spectral resolution and spectrophotometric
calibration, such that using the raw measurements may result in
systematic errors in evaluating spectral subtypes and the metallicity
$\zeta$ parameter. In Section 3.2 we outline a procedure for
calculating corrected indices, based on a calibration of systematic
offsets between two observatories. A proper calibration requires that
large numbers of stars be re-observed every time a new observatory
and/or instrumental setup is used, in order to calibrate the offsets
and correct the spectral indices. Only the corrected spectral indices
can be used reliably in the spectral subtype and $\zeta$
relationships, which are calibrated with respect to the corrected
values. We adopt the Palomar-MSU measurement as our standard of
reference for the spectral indices, and correct our MDM and UH
values accordingly.

In the end, this catalog provides a useful list of targets for
exoplanet searches, especially those based on the radial velocity
variation method. Current methods and instruments require relatively bright
stars to be efficient, and the stars presented in our spectroscopic
catalog all constitute targets of choice, having been vetted for
background source contamination. Our accurate spectral types will be
useful to guide radial velocity surveys and selct stars of
comparatively lower masses.

We also provide diagnostics for chromospheric activity from H$\alpha$
emission, X-ray flux excess, and UV excess. Besides being useful to
identify more challenging sources for radial velocity surveys, they
also isolate the younger stars in the census. Follow-up radial
velocity observations could tie some of the stars to nearby moving
groups, and these objects would be prime targets for exoplanet
searches with direct imaging methods.


\acknowledgments

{\bf Acknowledgments}

This material is based upon work supported by the National Science
Foundation under Grants No. AST 06-07757, AST 09-08419, and AST
09-08406.  We thank Greg Aldering for countless instances of
assistance with SNIFS, the telescope operators of the UH 2.2m
telescope, and Justin Troyer for observing assistance. We thank Bob
Barr and the staff at the MDM observatory for their always helpful
technical assistance.




\end{document}